\documentclass[lettersize,journal]{IEEEtran}
\usepackage{amsmath,amsfonts}
\usepackage{amssymb}
\usepackage{algorithm}
\usepackage{array}
\usepackage[font={footnotesize}]{caption}
\usepackage{subcaption}
\usepackage{textcomp}
\usepackage{stfloats}
\usepackage{url}
\usepackage{verbatim}
\usepackage{balance}
\usepackage{graphicx}
\usepackage{cite}
\usepackage{mathtools}
\usepackage[nolist]{acronym}
\usepackage{bm}
\usepackage[mode=build]{standalone}
\usepackage{tikz}
\usepackage{multirow}
\usepackage{pgfplots}
\usepackage{bigfoot} 
\usepackage[multiple]{footmisc}
\usetikzlibrary{arrows.meta}
\usetikzlibrary{spy}
\tikzset{every picture/.style={line width=0.6pt}}
\usetikzlibrary{arrows.meta,
                calc, 
                backgrounds,
                chains,
                fit,
                quotes,
                shapes.geometric,
                positioning,
                intersections,
                spath3}

\usepgfplotslibrary{colormaps} 
\usetikzlibrary{pgfplots.colormaps} 

\usepackage{algpseudocode}

\algdef{SE}[SUBALG]{Indent}{EndIndent}{}{\algorithmicend\ }%
\algtext*{Indent}
\algtext*{EndIndent}

\newtheorem{example}{Example}

\newcolumntype{C}[1]{ >{\centering\arraybackslash} m{#1} }

\usepackage{booktabs, ragged2e}

\usepackage[normalem]{ulem} 

\newcommand{\bmA}{\bm{A}}

\newcommand{\bmB}{\bm{B}}

\newcommand{\Exp}{\text{Exp}}
\newcommand{\bbold}{\bm{b}}
\newcommand{\complexset}[2]{ \mathbb{C}^{#1 \times #2}  }
\newcommand{\mathbbC}{\mathbb{C}}
\newcommand{\Ngrid}{N_{\rm{grid}}}
\newcommand{\Ntheta}{N_{\theta}}
\newcommand{\Ntau}{N_{\tau}}

\newcommand{\SSb}{\boldsymbol{S}}
\newcommand{\hermit}{\mathsf{H}}
\newcommand{\trpose}{\top}

\newcommand{\bma}{\bm{a}}

\newcommand{\aperturbed}{\bm{a}_{\rm{pert}}}
\newcommand{\conj}{ {\ast} }
\newcommand{\U}{ \mathcal{U} }
\newcommand{\cnormal}{ \mathcal{CN} }
\newcommand{\normal}{ \mathcal{N} }

\newcommand{\boldone}{{ {\boldsymbol{1}} }}

\newcommand{\Deltaf}{ \Delta_f }
\newcommand{\Tmax}{ T_{\rm{max}} }

\newcommand{\Yr}{ \bm{Y}_r }
\newcommand{\Yrtilde}{ \widetilde{\bm{Y}}_r }

\newcommand{\boldPhi}{ \mathbf{\Phi} }

\newcommand{\boldPsi}{ \mathbf{\Psi} }
\newcommand{\alphab}{ \boldsymbol{\alpha} }

\newcommand{\Tcp}{ T_{\rm{cp}} }

\newcommand{\sumt}{ \sum_{t=1}^T }
\newcommand{\bmf}{ \bm{f} }

\newcommand{\bmx}{ \bm{x} }
\newcommand{\bmm}{ \bm{m} }

\newcommand{\setM}{ \mathcal{M} }
\newcommand{\bmrho}{ \bm{\rho} }
\newcommand{\bmN}{ \bm{N} }
\newcommand{\That}{ \widehat{T} }
\newcommand{\yc}{ \bm{y_c} }

\newcommand{\bmn}{ \bm{n} }
\newcommand{\thetamax}{\theta_{\max}}
\newcommand{\thetamin}{\theta_{\min}}

\newcommand{\Rmax}{R_{\max}}
\newcommand{\Rmin}{ R_{\min}}

\newcommand{\alphahat}{\widehat{\alpha}}
\newcommand{\alphahatb}{\widehat{\alphab}}
\newcommand{\thetahat}{\widehat{\theta}}

\newcommand{\tauhat}{\widehat{\tau}}

\newcommand{\ihat}{\hat{i}}
\newcommand{\jhat}{\hat{j}}
\newcommand{\sigmalambda}{\sigma_{\lambda}}

\newcommand{\vecc}[1]{ {\rm{vec}}(#1)  }

\newcommand{\bmkappa}{\bm{\kappa}}
\newcommand{\mhat}{\widehat{m}}

\newcommand{\bmd}{\bm{d}}

\newcommand{\deltatheta}{\delta_{\theta}}
\newcommand{\deltaR}{\delta_{R}}

\newcommand{\thetagrid}{\bm{\theta}_{\rm{grid}}}
\newcommand{\taugrid}{\bm{\tau}_{\text{grid}}}

\newcommand{\sigmarcst}{{ \sigma_{{\rm{rcs}},t} }}
\newcommand{\sigmarcsmean}{{ \sigma_{{\rm{mean}}} }}

\newcommand{\psit}{ \widetilde{\psi} }
\newcommand{\taut}{ \widetilde{\tau} }
\newcommand{\thetat}{  \widetilde{\theta} }
\newcommand{\thetatildemin}{  \widetilde{\theta}_{1,\min} }
\newcommand{\thetatildemax}{  \widetilde{\theta}_{1,\max} }
\newcommand{\sigmarcstt}{{ \widetilde{\sigma}_{{\rm{rcs}},t} }}
\newcommand{\Ttilde}{\widetilde{T}}
\newcommand{\Rtilde}{ \widetilde{R} }

\newcommand{\bmtheta}{\bm{\theta}}

\newcommand{\bmvarthetainterval}{\bm{\vartheta}_{\rm{int}}}

\newcommand{\bmr}{\bm{r}}

\newcommand{\angledelaymap}{\bm{L}(\Yrtilde^{(I)})}
\newcommand{\admaptilde}{\widetilde{\bmL}}
\newcommand{\admapbar}{\widebar{\bm{L}}}

\newcommand{\realset}{\mathbb{R}}

\newcommand{\Pset}{\mathcal{P}}
\newcommand{\Psethat}{\hat{\mathcal{P}}}

\newcommand{\bmp}{\bm{p}}
\newcommand{\bmphat}{\hat{\bmp}}
\newcommand{\bmpant}{\bm{p}_{\text{ant}}}

\newcommand{\dcut}{d^{(\gamma)}}

\newcommand{\thetamean}{\theta_{\text{mean}}}

\newcommand{\Deltatheta}{\Delta_{\theta}}

\newcommand{\SNR}{\text{SNR}}

\newcommand{\degreee}{^{\circ}}

\newcommand{\bmL}{\bm{L}}
\newcommand{\bmu}{\bm{u}}
\newcommand{\bmuhat}{\hat{\bm{u}}}

\newcommand{\bmuenc}{\bmu_{\rm{enc}}}

\renewcommand{\Pr}{p}
\newcommand{\pmd}{\Pr_{\text{md}}}
\newcommand{\pfa}{\Pr_{\text{fa}}}
\newcommand{\Expectation}{\mathbb{E}}

\newcommand{\norm}[1]{\left\lVert#1\right\rVert}

\newcommand{\normb}[1]{\Big\lVert#1\Big\rVert}
\DeclarePairedDelimiter\abs{\lvert}{\rvert}%
\DeclarePairedDelimiter\absbigs{\big\lvert}{\big\rvert}%

\newcommand{\SNRr}{\SNR_{\mathrm{r}}}
\newcommand{\SNRc}{\SNR_{\mathrm{c}}}
\newcommand{\Tmaxtilde}{\widetilde{T}_{\max}}

\newcommand{\bmdgrid}{\bm{d}_{\mathrm{grid}}}
\newcommand{\Nd}{N_{\bm{d}}}
\newcommand{\bmJcal}{\bm{\mathcal{J}}}

\newcommand{\bmxi}{\bm{\xi}}
\newcommand{\bigO}{\mathcal{O}}

\newcommand{\iotatheta}{\iota_{\theta}}
\newcommand{\iotaR}{\iota_R}

\makeatletter
\newcommand*\rel@kern[1]{\kern#1\dimexpr\macc@kerna}
\newcommand*\widebar[1]{%
  \begingroup
  \def\mathaccent##1##2{%
    \rel@kern{0.8}%
    \overline{\rel@kern{-0.8}\macc@nucleus\rel@kern{0.2}}%
    \rel@kern{-0.2}%
  }%
  \macc@depth\@ne
  \let\math@bgroup\@empty \let\math@egroup\macc@set@skewchar
  \mathsurround\z@ \frozen@everymath{\mathgroup\macc@group\relax}%
  \macc@set@skewchar\relax
  \let\mathaccentV\macc@nested@a
  \macc@nested@a\relax111{#1}%
  \endgroup
}
\makeatother

\pgfplotsset{
  log x ticks with fixed point/.style={
      xticklabel={
        \pgfkeys{/pgf/fpu=true}
        \pgfmathparse{exp(\tick)}%
        \pgfmathprintnumber[fixed relative, precision=3]{\pgfmathresult}
        \pgfkeys{/pgf/fpu=false}
      }
  }
}

\begin{acronym}[ACRONYM]
\acro{6G}{6th generation wireless systems}

\acro{AE}{autoencoder}
\acro{AoA}{angle-of-arrival}
\acro{AoD}{angle-of-departure}

\acro{BCE}{binary cross-entropy}

\acro{CCE}{categorical cross-entropy}
\acro{CNN}{convolutional neural network}
\acro{CSI}{channel state information}

\acro{DL}{deep learning}
\acro{DFT}{discrete Fourier transform}

\acro{GOSPA}{generalized optimal sub-pattern assignment}

\acro{IRS}{intelligent reflecting surface}
\acro{ISTA}{iterative shrinkage-thresholding algorithm}
\acro{ISAC}{integrated sensing and communication}

\acro{JRC}{joint radar and communication}

\acro{LOS}{line-of-sight}
\acro{LS}{least-squares}

\acro{NLOS}{non-line-of-sight}

\acro{MB-ML}{model-based machine learning}
\acro{MIMO}{multiple-input multiple-output}
\acro{MISO}{multiple-input single-output}
\acro{ML}{machine learning} 
\acro{MLE}{maximum likelihood estimation} 
\acro{MLP}{multilayer perceptron}
\acro{MSE}{mean squared error}

\acro{NLL}{negative log-likelihood}
\acro{NN}{neural network}
\acro{NNBL}{neural-network-based learning}

\acro{OFDM}{orthogonal frequency-division multiplexing}
\acro{OMP}{orthogonal matching pursuit}
\acro{OSPA}{optimal sub-pattern assignment}

\acro{PDF}{probability density function}
\acro{PMF}{probability mass function}
\acro{PSD}{power spectral density}

\acro{QAM}{quadrature amplitude modulation}
\acro{QPSK}{quadrature phase shift keying}

\acro{ReLU}{Rectified Linear Unit}
\acro{RMSE}{root mean squared error}
\acro{ROC}{receiver operating characteristic}

\acro{SER}{symbol error rate}
\acro{SNR}{signal-to-noise ratio}

\acro{ULA}{uniform linear array}

\acro{V2I}{vehicle-to-infrastructure}
\end{acronym}

\begin{document}

\title{Model-Based End-to-End Learning for Multi-Target Integrated Sensing and Communication under Hardware Impairments}

\author{Jos\'{e} Miguel Mateos-Ramos,~\IEEEmembership{Student Member,~IEEE}, 
        Christian H\"{a}ger,~\IEEEmembership{Member,~IEEE},
        Musa Furkan Keskin,~\IEEEmembership{Member,~IEEE}, 
        Luc Le Magoarou,~\IEEEmembership{Member,~IEEE}, 
        Henk Wymeersch,~\IEEEmembership{Fellow,~IEEE}
\thanks{
This work was supported, in part, by a grant from the Chalmers AI Research Center Consortium (CHAIR), by the National Academic Infrastructure for Supercomputing in Sweden (NAISS), the Swedish Foundation for Strategic Research (SSF) (grant FUS21-0004, SAICOM), Hexa-X-II, part of the European Union’s Horizon Europe research and innovation programme under Grant Agreement No 101095759, and Swedish Research Council (VR grant 2022-03007). The work of C.~Häger was also supported by the Swedish Research Council under grant no. 2020-04718.}
\thanks{Jos\'{e} Miguel Mateos-Ramos, Christian H\"{a}ger,  Musa Furkan Keskin and Henk Wymeersch are with the Department of Electrical Engineering, Chalmers University of Technology, Sweden (email: josemi@chalmers.se; christian.haeger@chalmers.se; furkan@chalmers.se; henkw@chalmers.se).}%
\thanks{Luc Le Magoarou is with INSA Rennes, CNRS, IETR - UMR 6164, F-35000, Rennes, France (email: Luc.Le-Magoarou@insa-rennes.fr).}
}

\markboth{IEEE Transactions on Wireless Communications,~Vol.~X, No.~Y, Month~Year}%
{Shell \MakeLowercase{\textit{et al.}}: A Sample Article Using IEEEtran.cls for IEEE Journals}


\maketitle

\begin{abstract}
    We study model-based end-to-end learning in the context of integrated sensing and communication (ISAC) under hardware impairments. 
    Hardware impairments are usually addressed by means of array calibration with a focus on communication performance. However, residual impairments may exist that affect sensing performance. This paper proposes a data-driven framework for mitigating such impairments.
    A monostatic orthogonal frequency-division multiplexing (OFDM) sensing and multiple-input single-output (MISO) communication scenario is considered, incorporating hardware imperfections at the ISAC transceiver antenna array. 
    Since conventional ISAC signal processing algorithms rely on mathematical models of the wireless channel, a mismatch occurs between the assumed mathematical models and the underlying reality in the presence of hardware impairments. We first study the detrimental effects of such impairments at the transmitter and receiver side of the proposed scenario, showcasing different levels of degradation on communication and sensing performances.
    As the core contribution of this work, we propose a novel differentiable version of the orthogonal matching pursuit (OMP) algorithm that is suitable for multi-target sensing and allows for efficient end-to-end learning of the hardware impairments. 
    Based on the differentiable OMP, we devise two model-based parameterization strategies of the ISAC beamformer and sensing receiver to account for hardware impairments: (i) learning a dictionary of steering vectors for different angles and (ii) learning the parameterized hardware impairments.
    We carry out a comprehensive performance analysis of the proposed model-based learning approaches and a strong baseline consisting of least-squares beamforming, conventional OMP, and maximum-likelihood symbol detection for communication.
    Results show that by parameterizing the hardware impairments, learning approaches offer gains in terms of higher detection probability, position estimation accuracy, and lower symbol error rate (SER) compared to the baseline. We demonstrate that learning the parameterized hardware impairments outperforms learning a dictionary of steering vectors, also exhibiting the lowest complexity.
\end{abstract}
\vspace{-0.3cm}
\section{Introduction} \label{sec:introduction}
\IEEEPARstart{N}{ext}-generation wireless communication systems are expected to operate at higher carrier frequencies to meet the data rate requirements necessary for emerging use cases such as smart cities, e-health, and digital twins for manufacturing \cite{matthaiou2021road, 6g_vision_2023}. Higher carrier frequencies also enable new functionalities, such as \ac{ISAC}. \ac{ISAC} aims to integrate radar and communication capabilities in one joint system, which enables hardware sharing, energy savings, and improved channel estimation via sensing-assisted communications, among other advantages \cite{tan2021integrated, wymeersch2021integration, liu2022integrated}. 
\Ac{ISAC} has been mainly considered by means of dual-functional waveforms. For instance, radar signals have been used for communication \cite{lampel2019performance, lazaro2021car2car}, while communication waveforms have proven to yield radar-like capabilities \cite{liu2022integrated, zhang2021enabling}. Furthermore, optimization of waveforms to perform both tasks simultaneously has also been studied \cite{johnston2022mimo, chen2021joint, OFDM_DFRC_TSP_2021}.
However, conventional ISAC approaches degrade in performance under model mismatch, i.e., if the underlying reality does not match the assumed mathematical models. In particular at high carrier frequencies, hardware impairments can severely affect the system performance and hardware design becomes very challenging \cite{chowdhury20206g, jiang2021road}. This increases the likelihood of model mismatch in standard approaches, and problems become increasingly difficult to solve analytically if hardware impairments are considered.

\IEEEpubidadjcol 

\Ac{DL} approaches based on large \acp{NN} have proven to be useful under model mismatch or complex optimization problems \cite{mason2017deep, demir2019channel}. \Ac{DL} does not require any knowledge about the underlying models as it is optimized based on training data, which inherently captures the potential impairments of the system. 
\Ac{DL} has been investigated in the context of \ac{ISAC} for a vast range of applications, such as predictive beamforming in vehicular networks \cite{mu2021integrated, liu2022learning},  multi-target sensing and communication in THz transmissions \cite{wu2022sensing}, or efficient resource management \cite{yao2022joint, wang2022reinforcement}.
However, most previous works on DL for ISAC consider single-component optimization, either at the transmitter or receiver. On the other hand, end-to-end learning \cite{o2017introduction} of both the transmitter and receiver has proven to enhance the final performance of radar \cite{jiang2021joint} and communication \cite{aoudia2021end} systems. 
End-to-end learning in ISAC was applied 
to perform single-target angle estimation under hardware impairments \cite{mateos2022end} and multi-snapshot angle estimation \cite{muth2023autoencoder}.
Nevertheless, \ac{DL} approaches often lack interpretability and require large amounts of training data to obtain satisfactory performance. 

To overcome the disadvantages of large \ac{DL} models, \textit{\ac{MB-ML}} \cite{shlezingermodel} instead parameterizes existing models, designs, and algorithms while maintaining their overall computation graph as a blueprint.
This allows initializing trainable parameters at an already good starting point, requiring less training data to optimize, and typically also offers a better understanding of the learned parameters. 
A popular example of \ac{MB-ML} learning is \textit{deep unfolding} \cite{Hershey2014,unrolling_spmag_2021}, where iterative algorithms are ``unrolled'' and interpreted as multi-layer computation graphs. 
In the context of {radar-like} sensing, 
deep unfolding has been applied in \cite{xiao2020deepfpc, wu2022doa} to angle estimation, showing enhanced accuracy with respect to \ac{DL} and begin able to compensate for array imperfections, respectively.
Related to communications, deep unfolding has been applied to massive \ac{MIMO} channel estimation in \cite{yassine2022mpnet}, where classical steering vector models are used as a starting point and then optimized to learn the system hardware impairments, by unfolding the matching pursuit {(MP)} algorithm \cite{Mallat1993}. This approach was later refined to reduce the required number of learnable parameters in \cite{Chatelier2022}. 
Previous \ac{MB-ML} approaches {in the context of radar-like sensing \cite{xiao2020deepfpc, wu2022doa} and communications \cite{yassine2022mpnet, Chatelier2022}} exhibit three primary shortcomings that can limit their effectiveness in practical scenarios. 
Firstly, they {all} focus on receiver learning; however, end-to-end learning of transmitter and receiver, which holds great potential given its promising performance in model-free \ac{DL} applications \cite{jiang2021joint, aoudia2021end}, remains unexplored. 
Secondly, sensing works \cite{xiao2020deepfpc, wu2022doa} only investigate angle estimation, which means that multi-target positioning has not been studied before. 
{In addition, the communication works in \cite{yassine2022mpnet,Chatelier2022} consider estimation of communication channels without extracting geometric path parameters.}
Finally, while \ac{MB-ML} has been utilized to address challenges related to {either} sensing {\cite{xiao2020deepfpc, wu2022doa} {or} communications \cite{yassine2022mpnet,Chatelier2022}}, the trade-offs between sensing and communication both at training and testing stages have not been investigated yet.
\begin{figure}[tb]
    \centering
    \includegraphics[width=0.4\textwidth]{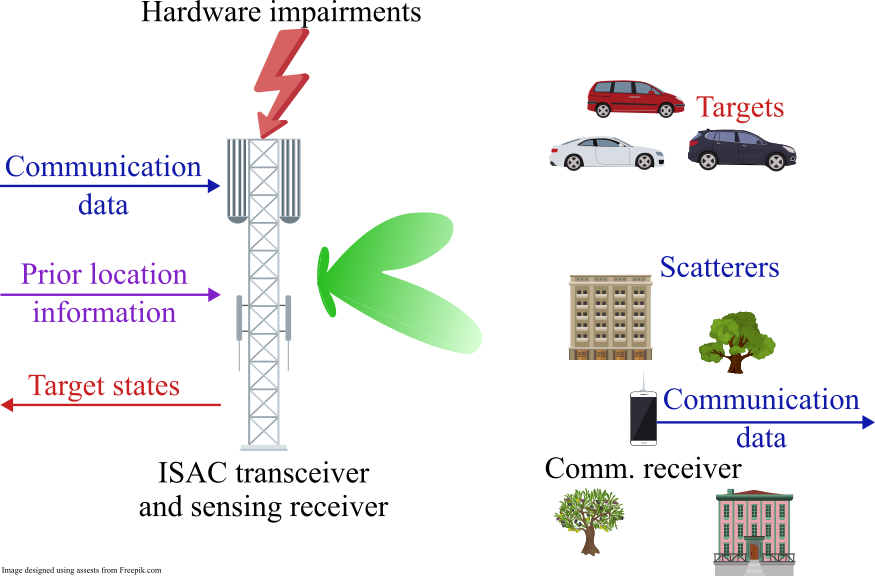}
    \caption{Considered scenario, where an impaired multi-antenna ISAC transmitter is optimized based on prior information of the location of the targets and the communication receiver. The co-located sensing receiver estimates the targets' states (target presence probability and position), while the single-antenna communication receiver retrieves the transmitted communication data.}
    \label{fig:general}
\end{figure} 
This paper {studies} end-to-end \ac{MB-ML} for \ac{ISAC}, focusing on the effect of hardware impairments in the ISAC transceiver. 
{We consider} a \ac{MIMO} monostatic sensing and \ac{MISO} communication scenario (as depicted in Fig.~\ref{fig:general}).
Our main contribution in this work is to propose an end-to-end MB-ML approach to learn the unknown hardware impairments. To this end, we require that the transmit and receive operations are: (i) differentiable, to enable gradient backpropagation and (ii) parameterizable by the hardware impairments. To achieve these requirements, we ground our proposed MB-ML approach in a model-based baseline consisting of a \ac{LS} beamformer at the transmitter and an \ac{OMP} position estimator at the co-located sensing receiver. 
After training, the learned impairments are used by the model-based baseline instead of the initially assumed ideal model parameters. 

{
This work significantly extends our preliminary work in \cite{mateos2022model} and other recent works on  
 MB-ML 
\cite{xiao2020deepfpc, wu2022doa, yassine2022mpnet, Chatelier2022} and end-to-end learning for ISAC \cite{mateos2022end, muth2023autoencoder}
(see Tab.~\ref{tab:comparison} for an overview) by addressing the three shortcomings mentioned above: 

\begin{itemize}
    \item \textbf{End-to-end learning:} Compared to prior related works on MB-ML \cite{xiao2020deepfpc, wu2022doa, yassine2022mpnet, Chatelier2022}, which focus on receiver learning, we consider simultaneous transmitter and receiver, i.e., end-to-end learning. 
    Our MB-ML approach parameterizes and optimizes both the ISAC beamformer and sensing receiver, allowing for end-to-end learning of hardware impairments. 
    
    \item \textbf{Joint angle--range processing: }
    We consider orthogonal frequency-division multiplexing (OFDM) transmission, which enables joint angle and range (and, hence, position) estimation. 
    Thus, compared to our previous work \cite{mateos2022model}, as well as the recent works in \cite{mateos2022end, muth2023autoencoder} and 
\cite{xiao2020deepfpc, wu2022doa}, which all focus on angle estimation, in this work we also estimate the range of targets in the scene, significantly augmenting the proposed end-to-end learning approach. 
Although \cite{Chatelier2022} does explore OFDM, it is confined to the single-input single-output (SISO) scenario.    
    \item \textbf{MB-ML for ISAC: }
    While MB-ML has been applied to either sensing \cite{xiao2020deepfpc, wu2022doa} or communication \cite{yassine2022mpnet, Chatelier2022} scenarios separately, our approach is designed to address the unique challenges of ISAC systems, such as \textit{(i)} 
    handle hardware impairments at the ISAC transmit array for effective ISAC beamforming (considering the trade-offs between the two functionalities), and \textit{(ii)} mitigating the detrimental impacts of hardware impairments on the detection and positioning performance of multiple targets.
    This work significantly expands on the conference version \cite{mateos2022model}, which (besides the already mentioned extension to range estimation) was limited to single-target scenarios. 
\end{itemize}}

In the rest of this paper, we first describe the mathematical \ac{ISAC} system model in Sec.~\ref{sec:model}. Then, we describe the two approaches to perform target positioning and communication:  
the baseline in Sec.~\ref{sec:baseline} and \ac{MB-ML} in Sec.~\ref{sec:md_learning}. The main \ac{ISAC} results are presented and discussed in Sec.~\ref{sec:results} before the concluding remarks of Sec.~\ref{sec:conclusions}.

\textit{Notation.} We denote column vectors as bold-faced lower-case letters, $\bm{a}$, and matrices as bold-faced upper-case letters, $\bm{A}$. A column vector whose entries are all equal to 1 is denoted as $\bm{1}$. The identity matrix of size $N\times N$ is denoted as $\bm{I}_N$. The transpose and conjugate transpose operations are denoted by $(\cdot)^\trpose$ and $(\cdot)^\hermit$, respectively. The $i$-th element of a vector and the $(i,j)$-th element of a matrix are denoted by $[\bm{a}]_i$ and $[\bmA]_{i,j}$. The element-wise product between two matrices is denoted by $\bmA\odot\bmB$ and $\otimes$ denotes the Kronecker product. $\vecc{\cdot}$ denotes matrix vectorization operator. Sets of elements are enclosed by curly brackets and intervals are enclosed by square brackets. The set $\{x\in\realset|x\geq0\}$ is denoted as $\realset_{\geq0}$. The cardinality of a set $\mathcal{X}$ is denoted by $\abs{\mathcal{X}}$. 
The uniform, circularly-symmetric complex Gaussian, and exponential distributions are denoted as $\U$, $\cnormal$, and $\Exp$, respectively. 
The Euclidean vector norm is represented by $\lVert\cdot\rVert_2$.
The indicator function is denoted by $\mathbb{I}\{\cdot\}$. The greatest integer less than or equal to $x$ is denoted as $\lfloor x \rfloor$.
\vspace{-0.4cm}
\begin{table*}[t]
\centering%
\caption{{Comparison between this and closely related prior work. (FPCA: fixed-point continuation algorithm, ISTA: iterative shrinkage-thresholding algorithm, MP: matching pursuit, OMP: orthogonal matching pursuit, MAP: maximum a posteriori, RX: receiver)}}%
{%
\begin{tabular}{c||C{3.8cm}|C{2.7cm}|C{3.8cm}|C{2.5cm}|C{2.5cm}}%
\toprule%
  Ref.   & ISAC & end-to-end & model-based & multi-target & target position  \\
\midrule
\midrule
\cite{mateos2022end} & yes & yes & no & no & no (angle only) \\
\cite{muth2023autoencoder} & yes & yes & no & yes & no (angle only) \\
\cite{xiao2020deepfpc} & no (sensing only) & no (RX only) & yes (unfolded FPCA) & yes & no (angle only) \\
\cite{wu2022doa} & no (sensing only) & no (RX only) & yes (unfolded ISTA) & yes & no (angle only) \\
\cite{yassine2022mpnet} & no (comm.~channel est. only) & no (RX only) & yes (unfolded MP) & N/A$^\dagger$ & N/A$^\dagger$ \\
\cite{Chatelier2022} & no (comm.~channel est. only) & no (RX only) & yes (unfolded MP) & N/A$^\dagger$ & N/A$^\dagger$ \\
\cite{mateos2022model}$^*$ & yes & yes & yes (param.~MAP ratio test) & no & no (angle only) \\
\midrule
this work & yes & yes & yes (unfolded OMP) & yes & yes \\
\bottomrule%
\end{tabular}%
\RaggedRight

\smallskip

$\quad^{*}$conference version of this paper

$\quad^\dagger$not applicable: no geometric path parameters are extracted from the channel estimates

}%
\label{tab:comparison}
\end{table*}
\vspace{-0.25cm}
\section{System Model} \label{sec:model}
This section provides the mathematical models for the received sensing and communication signals, the \ac{ISAC} transmitted signal and the hardware impairments. In Fig.~\ref{fig:system_model}, a block diagram of the considered ISAC system is depicted.
\vspace{-0.45cm}
\subsection{Multi-target MIMO Sensing} \label{subsec:multi_target_model}
    We consider an ISAC transceiver consisting of an ISAC transmitter and a sensing receiver sharing the same \ac{ULA} of $K$ antennas, as shown in Fig.~\ref{fig:general}. 
    The transmitted signal consists of an  \ac{OFDM} waveform across $S$ {contiguous} subcarriers, with an inter-carrier spacing of $\Deltaf$ Hz. In the sensing channel, we consider at most $\Tmax$ possible targets. Then, the backscattered signal impinging onto the sensing receiver can be expressed over antenna elements and subcarriers as \cite{5G_NR_JRC_analysis_JSAC_2022,MIMO_OFDM_ICI_JSTSP_2021}
    \begin{align}
        \Yr = \sumt \psi_t \bma(\theta_t) \bma^\trpose(\theta_t) \bmf [\bmx(\bmm) \odot \bmrho(\tau_t)]^\trpose + \bm{W}, \label{eq:Z_r}
    \end{align}
    where $\Yr \in \complexset{K}{S}$ collects the observations in the spatial-frequency domains, $T \sim \U\{0,...,\Tmax\}$ is the instantaneous number of targets in the scene{\footnote{\label{fn_radar_los}{In \eqref{eq:Z_r}, we consider only \ac{LOS} reflections from nearby targets. While reflections from the ground or surrounding buildings can cause multiple echoes from each physical object, the highly directional nature of transmissions at mmWave frequencies, coupled with severe path loss, can significantly reduce the impact of such reflections, effectively leaving only \ac{LOS} echoes in the radar channel \cite{ieee802_isac_2020,roadAhead_JRC_2020}.}}}, 
    and ${\psi_t}$ represents the complex channel gain of the $t$-th target, {which involves the impact of path loss and radar cross section (RCS). The amplitude of $\psi_t$ is given by the radar range equation \cite[Eq.~(2.8)]{richards2005fundamentals}
    \begin{align}\label{eq_ch_gain}
        \lvert \psi_t \rvert^2 =  \frac{ \sigmarcst \lambda^2 }{ (4 \pi)^3 R_t^4 }  \,,
    \end{align}
    while its phase is uniformly distributed over $[0, 2\pi)$. In \eqref{eq_ch_gain}, $\sigmarcst$ is the RCS of the $t$-th target, which follows a Swerling model 3 \cite[Eq.~(2.2.6)]{richards2005fundamentals}), i.e., $\sigmarcst\sim\Exp(1/\sigmarcsmean)$, $\lambda = c/f_c$ is the carrier wavelength with $c$ and $f_c$ denoting the speed of light and the carrier frequency, respectively, and $R_t$ represents the range of the $t$-th target.}
    
    {In \eqref{eq:Z_r},} the steering vector of the ISAC transceiver \ac{ULA}\footnote{{The steering vector $\bma(\theta)$ is assumed to be frequency-independent (i.e., wideband effects can be ignored) since $\frac{S \Deltaf}{f_c} \frac{d}{\lambda} K = 0.0328 \ll 1$ for the simulation setting considered in Sec.~\ref{subsec:simulation_parameters} \cite{wideband_TSP_2018}.}} for an angular direction $\theta$ is, under no hardware impairments, $[\bma(\theta)]_k= \exp(-\jmath 2 \pi (k-(K-1)/2) d \sin (\theta ) / \lambda)$, $k=0,...,K-1$, with $d = \lambda / 2$. 
    The precoder $\bmf \in \mathbb{C}^K$ determines the transmit antenna radiation pattern, where $\norm{\bmf}^2=P$ is the transmit power. 
    Target ranges are conveyed by $\bmrho(\tau_t) \in \mathbb{C}^S$, with $[\bmrho(\tau_t)]_s = \exp(-j2\pi s \Deltaf\tau_t)$, $s=0,...,S-1$, and where $\tau_t = 2R_t/c$ represents the round-trip time of the $t$-th target. {To ensure the validity of \eqref{eq:Z_r}, the cyclic prefix duration $\Tcp$ of the OFDM waveform is taken to be larger than the round-trip delay of the furthermost target, i.e., $\Tcp \geq \tau_t \, \forall t$ \cite{Firat_OFDM_2012,MIMO_OFDM_ICI_JSTSP_2021}.} Moreover, the communication symbol vector $\bmx(\bmm) \in \mathbb{C}^S$ conveys a vector of messages $\bmm\in \setM^S$, each uniformly distributed from a set of possible messages $\setM$ {and satisfying $\mathbb{E}[\norm{\bmx(\bmm)}^2] = S$}. Finally, the receiver noise is represented by $\bm{W}$, with $[\bm{W}]_{i,j} \sim \cnormal(0,N_0 {S \Deltaf})$, {where $N_0$ denotes the noise \ac{PSD}}. Note that if $T=0$, only noise is received. From the complex channel gain and the noise, we define the {maximum achievable\footnote{\label{fn_sensing_snr}{Note that the actual SNR depends on $|\bma^{\top}(\theta)\bmf|^2 \leq PK$, which in turn depends on the algorithm to compute $\bmf$. Maximum achievable SNR refers to using $PK$ as an upper bound for $|\bma^{\top}(\theta)\bmf|^2$.}} sensing \ac{SNR} for a given target as $\SNR_r = PK\Expectation[|\psi_t|^2]/(N_0S\Delta_f)$.}

    The angles and ranges of the targets are uniformly distributed within an uncertainty region, i.e., $\theta_t \sim \U[\thetamin, \thetamax]$ and $R_t \sim \U[\Rmin, \Rmax]$. However, uncertainty regions might change at each new transmission. The position of each target is computed from target angle $\theta_t$ and range $R_t$ as $\bmp_t = [R_t\cos{(\theta_t)}, R_t\sin{(\theta_t)}]^\top$.
    The transmitter and the sensing receiver are assumed to have knowledge of a prior estimate of the targets location, i.e., $\{\thetamin, \thetamax, \Rmin, \Rmax\}$, as in \cite{liu2022learning, ren2023fundamental}. In the considered monostatic sensing setup, the receiver has access to communication data $\bmx(\bmm)$, which enables removing its impact on the received {signal \eqref{eq:Z_r} via matched filtering\cite{OFDM_Radar_Corr_TAES_2020,reciprocalFilter_OFDM_2023}. Assuming \ac{QPSK} data (which is utilized for the simulations in Sec.~\ref{sec_perf_metric}), the received signal after matched filtering is
    \begin{align}
        \Yrtilde = \Yr \odot \boldone \bmx^\hermit(\bmm) = \sumt \alpha_t \bma(\theta_t) \bmrho^\trpose(\tau_t) + \bmN, \label{eq_Z_r2}
    \end{align}
    where $\alpha_t= \psi_t \bma^\trpose(\theta_t) \bmf $ and $\bmN = \bm{W} \odot \boldone \bmx^\hermit(\bmm)$}. With \ac{QPSK} data, {matched filtering does not lead to any performance loss for sensing \cite{holisticISAC_2024}.}\footnote{When using \acl{QAM} data, alternative approaches, such as the linear minimum mean-squared-error estimator in \cite{holisticISAC_2024}, can be employed {to mitigate the effect of the communication data on the received sensing signal, dealing with the changes in noise characteristics resulting from the element-wise product.}}
    
    The goal of the sensing receiver is to estimate the presence probability of each target in the scene, denoted as $\hat{\bm{u}} \in [0,1]^{\Tmax}$, which is later thresholded to provide a hard estimate of the target presence, $\hat{\bm{t}} \in \{0,1\}^{\Tmax}$. For all detected targets, the sensing receiver estimates their angles, $\hat{\bm{\theta}} \in [-\pi/2, \pi/2]^{\Tmax}$ and their ranges, $\hat{\bm{R}} \in \realset_{\geq 0}^{\Tmax}$, from which target positions are estimated. 
    
    \subsection{MISO Communication} 
    
    \begin{figure*}[t]
        \centering
        \begin{tikzpicture}[font=\scriptsize, >=stealth, blkCH/.style={draw,minimum height=0.8cm,text width=2.2cm, text centered},  blk/.style={draw,minimum height=0.8cm,text width=1.4cm, text centered}, blkNN/.style={fill=green!20, draw=green!30!black,minimum height=0.8cm,text width=1.4cm, text centered},  x=0.6cm,y=0.55cm]

\tikzset{threshold_fig/.pic={
\draw (-1,0)--(1,0);
\draw[red,line width=1.0 pt] (-1,-1)--(0,-1)--(0,1)--(1,1);
}}

\tikzset{
  bridging path/.initial=arc,
  bridging span/.initial=8pt,
  bridging gap/.initial=4pt,
  bridge/.style 2 args={
    spath/split at intersections with={#1}{#2},
    spath/insert gaps after
    components={#1}{\pgfkeysvalueof{/tikz/bridging span}},
    spath/join components upright
    with={#1}{\pgfkeysvalueof{/tikz/bridging path}},
    spath/split at intersections with={#2}{#1},
    spath/insert gaps after
    components={#2}{\pgfkeysvalueof{/tikz/bridging gap}},
  }
}

\tikz[overlay] \path[spath/save=arc] (0,0) arc[radius=1cm, start
    angle=180, delta angle=-180];

\path		
    (0,0)coordinate[](m){} 
    node(Encoder)[blk, right=3.5 of m]{Encoder (M-QAM)}

    coordinate[below=3.5 of m](theta){}
    node(Radar_precoder)[blkNN] at (theta -| Encoder){Precoding}
    coordinate[below=2 of theta](varphi){}
    node(Comm_precoder)[blkNN]at (varphi -| Encoder){Precoding}

    node(Joint_precoder)[blk, minimum height=1.4cm, text width=1.2cm, right=3.5 of $(Radar_precoder)!0.5!(Comm_precoder)$]{$\bmf(\eta,\phi)$}

    coordinate(help_tx) at (Encoder-|Joint_precoder)
    node(Tx_waveform)[blk, right=1 of $(help_tx)!0.5!(Joint_precoder)$]{{Transmit \\ signal}}
    coordinate[right=.5 of Tx_waveform](mid_tx_ch)

    node(Radar_CH)[blkCH, below right=-1 and 8 of Encoder]{Sensing Channel \\ $\Yr \sim \Pr(\Yr|\bmf, \bmx)$}
    node(Comm_CH)[blkCH, below = 2 of Radar_CH]{Comm. Channel \\ $\yc \sim \Pr(\yc|\bmf, \bmx)$}
    coordinate[below left=0.5 and 0.5 of Radar_CH.west](temp_radar)
    coordinate[above left=0.5 and 1 of Comm_CH.west](temp_comm)

    node(Removal)[blkCH, text width=2.5cm, right=2.7 of Radar_CH]{{Matched filtering \\ $\Yrtilde = \Yr \odot \boldone \bmx^\hermit(\bmm)$}}
    node(Radar_estimator)[blkNN, minimum height=1.5cm, below right=0.4 and 3.75 of Removal.north]{Sensing Estimator}
    coordinate[below left=2.1 and 4 of Removal.east](input_radar)
    coordinate[above right=-0.2 and 4 of Radar_estimator](output_radar_1)
    coordinate[below right =-0.2 and 4 of Radar_estimator](output_radar_2)
    ($(output_radar_1)!0.5!(output_radar_2)$)coordinate(output_radar_3)
    node(Threshold)[draw, minimum height=0.8cm, text width=0.8cm, right=0 of output_radar_1]{} 
    coordinate[right=1 of Threshold](output_threshold)

    node(Comm_decoder)[blkNN] at (Comm_CH -| Removal){Comm. Decoder}
    coordinate[right=2.5 of Comm_decoder](output_comm)

    coordinate[below=1 of Comm_CH](csi)
    (csi-|Comm_decoder)coordinate(csi_2) 
    
;
\draw[->] (m)--node[above]{$\bmm \in \setM^S$}(Encoder);

\draw[->] (theta)--node[above]{$\{\thetamin, \thetamax\}$}(Radar_precoder);
\draw[->] (varphi)--node[above]{$\{\thetat_{1,\min}, \thetat_{1,\max}\}$}(Comm_precoder);

\draw[->] (Radar_precoder)--node[above]{$\bmf_r \in \mathbb{C}^K$}(Radar_precoder -| Joint_precoder.west);
\draw[->] (Comm_precoder)--node[above]{$\bmf_c \in \mathbb{C}^K$}(Comm_precoder -| Joint_precoder.west);

\draw[->] (Joint_precoder)-|node[below]{$\bmf \in \mathbb{C}^K$}(Tx_waveform); 
\draw[->] (Encoder)-|node[above]{$\bmx(\bmm) \in \mathbb{C}^S$}(Tx_waveform);
\draw[-] (Tx_waveform)--(mid_tx_ch);
\draw[->] (mid_tx_ch)|-(Radar_CH);
\draw[->] (mid_tx_ch)|-(Comm_CH);

\draw[->] (Radar_CH)--node[above]{$\Yr \in \complexset{K}{S}$}(Removal);
\draw[->] (Removal)--(Removal-|Radar_estimator.west);
\draw[->] (input_radar)--node[above]{$\{\thetamin, \thetamax, \Rmin, \Rmax\}$}(input_radar-|Radar_estimator.west);
\draw[->] (Radar_estimator.east |- Threshold)--node[above]{$\hat{\bm{u}} \in [0,1]^{\Tmax}$}(Threshold);
\pic[scale=0.5]() at (Threshold) {threshold_fig};
\draw[->] (Threshold)--node[above]{$\hat{\bm{t}}$}(output_threshold);
\draw[->] (Radar_estimator.east |- output_radar_2)--node[above]{$\hat{\bmr} \in \mathbb{R}_{\geq 0}^{\Tmax}$}(output_radar_2);
\draw[->] (Radar_estimator.east |- output_radar_3)--node[above]{$\hat{\bmtheta} \in [-\frac{\pi}{2}, \frac{\pi}{2}]^{\Tmax}$}(output_radar_3);

\draw[->] (Comm_CH)--node[above]{$\yc \in \mathbb{C}^S$}(Comm_decoder);
\draw[->] (Comm_decoder)--node[above]{$\hat{\bmm} \in \setM^S$}(output_comm);

\draw[-, dotted] (Comm_CH)--(csi);
\draw[-, dotted] (csi)--node[above]{$\bmkappa =  \sum_{t=1}^{\Ttilde} \psit_t \bma^\trpose(\thetat_t) \bmf  \bmrho(\taut_t)$}(csi_2);
\draw[->, dotted] (csi_2)--(Comm_decoder);

\end{tikzpicture}
        \caption{Block diagram of the ISAC system model. The colored blocks can be implemented following standard neural-network-based learning \cite{mateos2022end, muth2023autoencoder}, the baseline of Secs.~\ref{subsec:baseline_beamformer}, \ref{subsec:baseline_sensing_rx}, or model-based learning of Sec.~\ref{sec:md_learning}. The precoding block applies the same mapping function for sensing and communication. Note that the sensing estimator is co-located with the ISAC transmitter.}
        \label{fig:system_model}
    \end{figure*}
    \label{subsec:comm_model}
    We assume that the communication receiver is equipped with a single antenna element. In this setting, the received OFDM signal at the communication receiver in the frequency domain is given by
    {\begin{align}
        \yc =  \sum_{t=1}^{\Ttilde} \psit_t \bma^\trpose(\thetat_t) \bmf [\bmx(\bmm) \odot \bmrho(\taut_t)] + \bmn, \label{eq:y_c}
    \end{align}
    where $\psit_t$, $\thetat_t$, and $\taut_t$ denote the complex channel gain, angle-of-departure, and delay of the $t$-th path, respectively, and $\Ttilde$ is the number of paths.} Complex Gaussian noise $\bmn \sim \cnormal(\bm{0},{N_0} {S \Deltaf}\bm{I}_S)$ is added at the receiver side. {Since we consider a single-user scenario with a single data stream $\bmx(\bmm)$, we adopt a frequency-flat precoding model $\bmf$ (i.e., single-stream beamforming \cite{MIMO_OFDM_ICI_JSTSP_2021,5G_NR_JRC_analysis_JSAC_2022}).}

    {In \eqref{eq:y_c}, $t=1$ represents the \ac{LOS} path, while $t>1$ are the \ac{NLOS} paths due to scattering off the objects in the environment. Accordingly, the channel gains are given by \cite[Eq.~(45)]{Zohair_5G_errorBounds_TWC_2018}
    \begin{align} \label{eq:y_c_gain}
        \lvert \psit_t \rvert^2 = \begin{cases}
	\lambda^2/ (4 \pi \Rtilde_1)^2 ,&~ t = 1   \\
	\sigmarcstt \lambda^2 / [(4 \pi)^3 \Rtilde^2_{t,1} \Rtilde^2_{t,2}] ,&~ t > 1
	\end{cases} \,, 
    \end{align}
    where $ \Rtilde_1$ is the distance of the \ac{LOS} path, $\sigmarcstt$ denotes the RCS of the scatterer for the $t$-th path, and $\Rtilde_{t,1}$ and $\Rtilde_{t,2}$ are the distances between TX-scatterer and scatterer-RX.}

    The communication receiver is assumed to be always present at a random position, such that {$\thetat_1 \sim \U[\thetat_{1,\min}, \thetat_{1,\max}]$}. The transmitter has also knowledge of {$\{\thetat_{1,\min}, \thetat_{1,\max}\}$. The OFDM system is designed such that the cyclic prefix is greater than the delay spread of the channel, i.e., $\Tcp \geq (\max_t\{\Rtilde_{t,1} + \Rtilde_{t,2}\} - \Rtilde_1)/c$}. {Based on \eqref{eq:y_c},} the receiver is fed with the \ac{CSI} {(i.e., the frequency response over $S$ subcarriers)}, which is given by
    {\begin{align} \label{eq:kappa}
        \bmkappa = \sum_{t=1}^{\Ttilde} \psit_t \bma^\trpose(\thetat_t) \bmf  \bmrho(\taut_t) \in \mathbb{C}^S  \,,
    \end{align}    
    where \eqref{eq:y_c} takes the equivalent form $\yc = \bmkappa \odot \bmx(\bmm) + \bmn$. We assume that this CSI is known to the communication receiver, e.g., based on pilot transmissions.} {The communication \ac{SNR} at the $s$-th subcarrier is defined as $\SNR_s = \Expectation[|[\bmkappa]_s|^2]/(N_0S\Deltaf)$, and the average communication SNR as $\SNR_c = \sum_{s=1}^S \SNR_s/S$}. The goal of the receiver is to retrieve the communication messages $\bmm$ that were transmitted.

    \subsection{ISAC Transmitter} \label{subsec:isac_tx}
    ISAC scenarios require the use of a radar--communication beamformer to provide adjustable trade-offs between the two functionalities. Using the multi-beam approach from \cite{zhang2018multibeam}, we design the {digital} ISAC beamformer, based on a sensing precoder $\bmf_r \in \mathbb{C}^K$ and a communication precoder $\bmf_c\in \mathbb{C}^K$, as
    \begin{align} \label{eq:isac_precoder_bench}
        \bmf(\eta,\phi) = \sqrt{P}\frac{\sqrt{\eta} \bmf_r + \sqrt{1-\eta}e^{\jmath \phi }\bmf_c}{\Vert\sqrt{\eta} \bmf_r + \sqrt{1-\eta}e^{\jmath \phi }\bmf_c \Vert } ~,
    \end{align}
    where $P$ is the transmitted power, $\eta \in [0,1]$ is the ISAC trade-off parameter, and $\phi \in [0,2 \pi)$ is a phase ensuring coherency between multiple beams.
    By sweeping over $\eta$ and $\phi$, we can explore the ISAC trade-offs of the considered system. The sensing precoder $\bmf_r$ illuminates the angular sector of the targets, $\{\thetamin, \thetamax\}$, whereas the communication precoder $\bmf_c$ illuminates the angular sector of the communication receiver, ${\{\thetatildemin, \thetatildemax\}}$. In Secs.~\ref{subsec:baseline_beamformer} and \ref{subsec:md_beamformer}, we detail how $\bmf_r$ and $\bmf_c$ are computed for the baseline and  \ac{MB-ML}, respectively. The same precoding function (with different inputs) is applied for sensing and communication, as represented in Fig.~\ref{fig:system_model}.

    \subsection{Hardware Impairments} \label{subsec:hardware_impairments}
    We study the effect of hardware impairments in the \ac{ULA} of the \ac{ISAC} transceiver, which affect the steering vectors of \eqref{eq:Z_r}, \eqref{eq_Z_r2}, \eqref{eq:y_c}. Impairments in the antenna array include mutual coupling, array gain errors, or antenna displacement errors, among others \cite{schenk2008rf}. Following the impairment models of \cite{chen2023modeling}, we consider two types of impairments: 
    \begin{enumerate}
        \item \emph{Unstructured impairments:} In this case, the true steering vector $\aperturbed(\theta)$ is unknown for all angles $\theta$, while the methods for beamforming design and signal processing assume the nominal steering vector  $\bma(\theta)$. If we consider a grid of possible angles with $N_{\theta}$ points, then the steering vectors require $K\times N_{\theta}$ complex values to be described. 
        \item \emph{Structured impairments:} In this case, the steering vector model is known, conditional on an unknown perturbation vector $\bmd$. We can thus write $\aperturbed(\theta;\bmd)$, where the meaning and dimensionality of $\bmd$ depend on the type of impairment.  In contrast to the unstructured impairments, the impairments are often described with a low-dimensional vector, independent of $N_{\theta}$. 
    \end{enumerate}

\begin{figure}[tb]
        \centering 
        \begin{subfigure}{0.45\textwidth}
        \includegraphics[width=\textwidth]{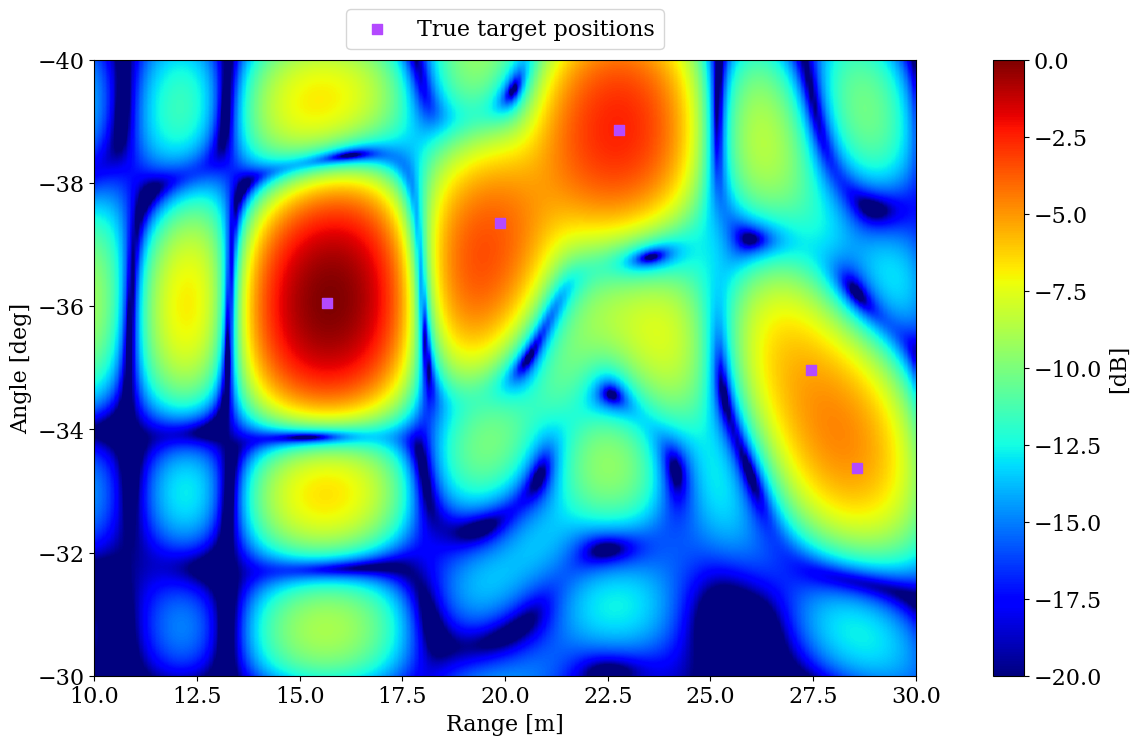}
        \caption{Ideal conditions}
        \end{subfigure}
        \begin{subfigure}{0.45\textwidth}
        \includegraphics[width=\textwidth]{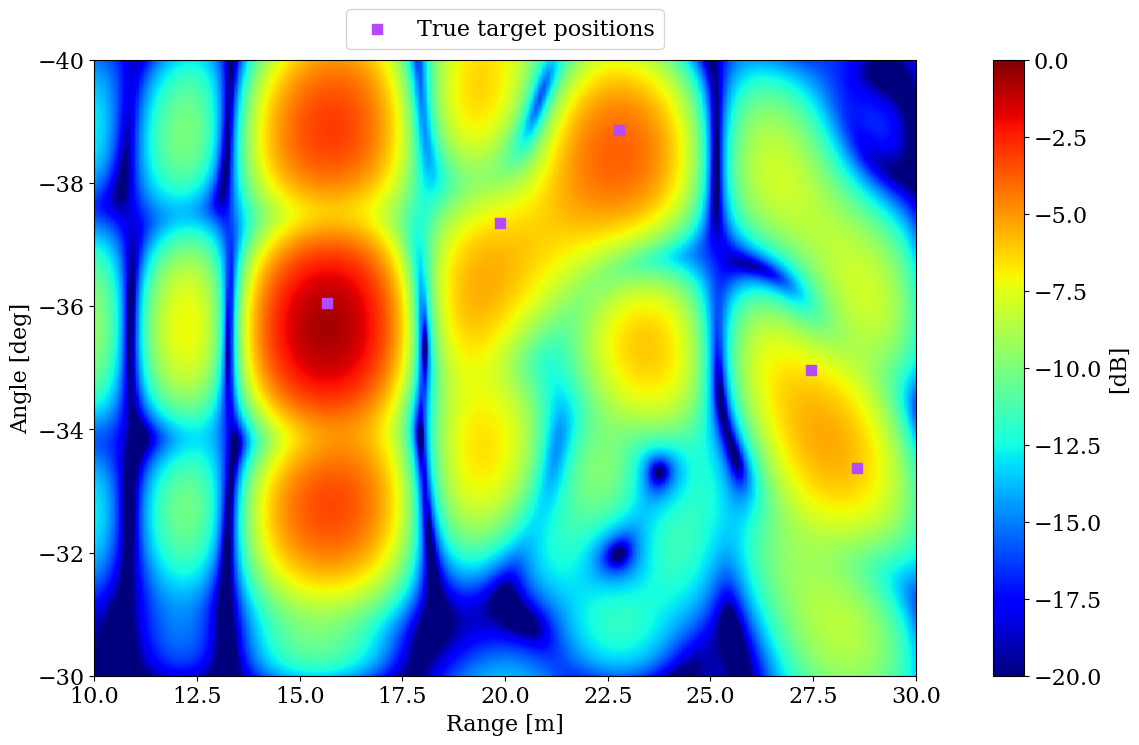}
        \caption{Hardware impairments}
        \end{subfigure}
        \caption{Example of the angle-delay map when $T = 4$ targets are present, under ideal conditions (top) and hardware impairments in the form of inter-antenna spacing mismatch (bottom), for {a sensing SNR of 6.5 dB}. Both maps are normalized with respect to the maximum value of the angle-delay map under ideal conditions. More details about the specific simulation parameters can be found in Sec.~\ref{subsec:simulation_parameters}.}
        \label{fig:angle_range_map}
    \end{figure}

    {
    \begin{example}[Impact of structured impairments at the transmitter] \label{ex_beamforming}
    Consider the example of inter-antenna spacing errors, where $\bmd \in {\mathbb{R}^K}$ and  $[\aperturbed(\theta; \bmd)]_k = \exp(-\jmath 2 \pi (k-(K-1)/2) [\bmd]_k \sin (\theta ) / \lambda)$, $k=0,...,K-1$. In Fig.~\ref{fig:example_bf}, the precoder response $|\bma(\vartheta)^\top \bmf|^2$ is shown for $\vartheta\in [-90\degreee, 90\degreee]$. The sensing and communication precoders $\bmf_r$ and $\bmf_c$ are designed to illuminate the sensing angular sector $[\thetamin, \thetamax] = [-40\degreee, -20\degreee]$ and the communication angular sector ${[\thetatildemin, \thetatildemax]} = [30\degreee, 40\degreee]$, respectively (further details on how to compute both precoders can be found in Sec.~\ref{subsec:baseline_beamformer}). When hardware impairments are introduced, the antenna power is spread into wider angular regions, reducing the effective power in the desired sensing and communication regions compared to the ideal precoder. This reduces the \ac{SNR} at both sensing and communication receivers, which degrades position estimation and message decoding performance.
    \end{example}
    \begin{figure}[tb]
        \centering         
        \includegraphics[width=0.47\textwidth]{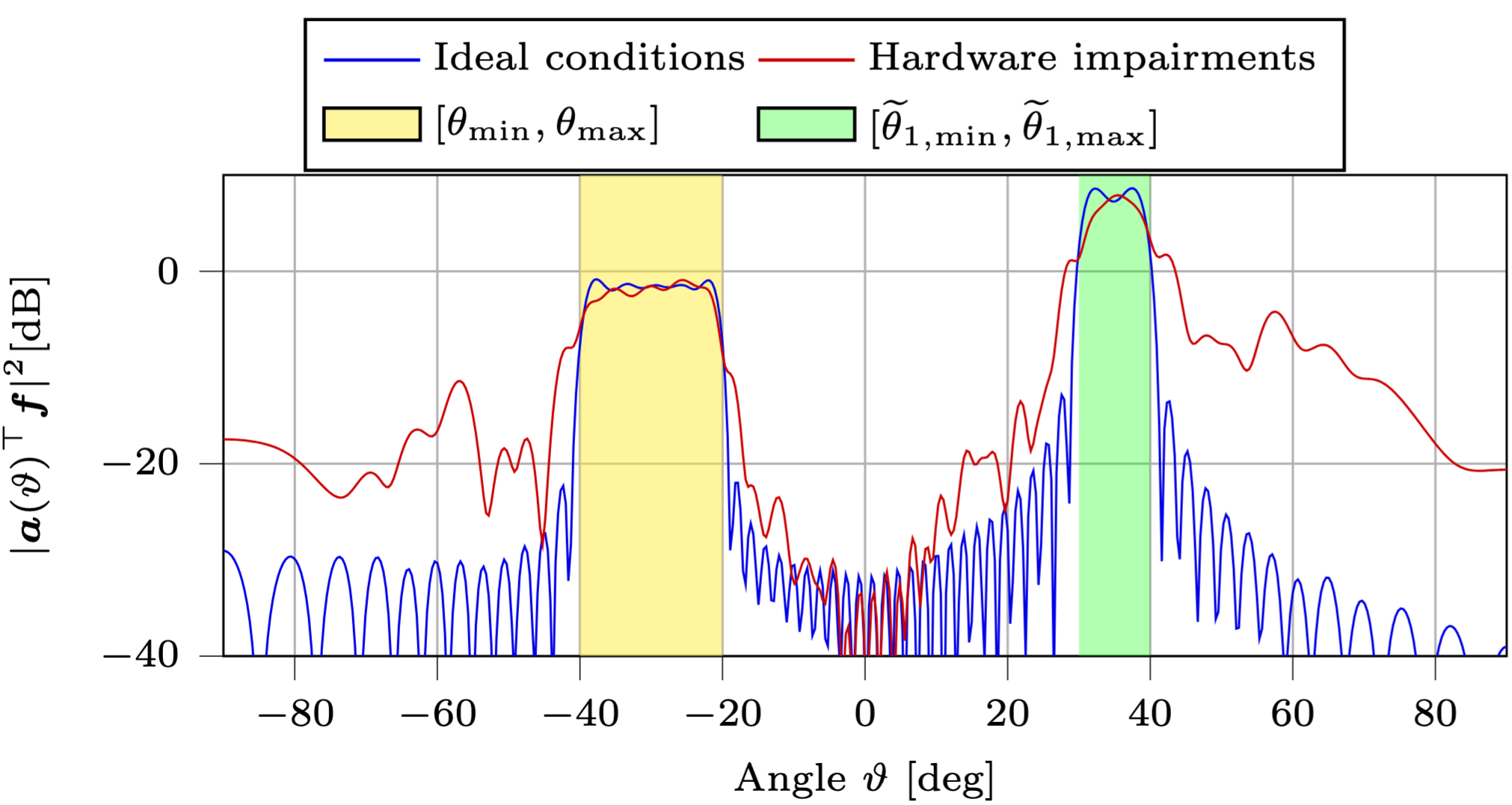}
        \caption{{Precoder response $|\bma(\vartheta)^\top \bmf|^2$ under ideal conditions and hardware impairments in the form of inter-antenna spacing mismatch, for a sensing angular sector $[\thetamin, \thetamax] = [-40\degreee, -20\degreee]$, and a communication angular sector $[{\thetatildemin, \thetatildemax}] = [30\degreee, 40\degreee]$. The parameters $P, \eta$ and $\phi$ in \eqref{eq:isac_precoder_bench} are $P=1$, $\eta=0.2$, and $\phi=0$. Both curves are normalized to integrate to the same value for values $\vartheta\in[-180\degreee, 180\degreee]$.}}
        \label{fig:example_bf}
    \end{figure}   
    }
    \begin{example}[Impact of structured impairments {at the sensing receiver}] \label{ex_impairments}
    Consider {the same kind of impairments as in Example~1.}
    In Fig.~\ref{fig:angle_range_map}, the angle-delay map (to be formally defined in Sec.~\ref{subsec:baseline_sensing_rx}) 
    is depicted under ideal conditions (top) and hardware impairments (bottom), when $T = 5$ targets are present. The main effect of hardware impairments is to create spurious lobes in the angular domain. 
    {As illustrated in Fig.~\ref{fig:angle_range_map}, these impairments can degrade angle estimation accuracy for targets, particularly those at 16 m and 20 m distances. Furthermore, the appearance of spurious lobes in the angular domain, especially from the target located at 16 m, may hinder the detection of nearby targets, complicating the detection process in areas close to this specific target.}    
    Another effect of hardware impairments is that the magnitude of the target lobes is decreased, which makes them harder to differentiate from noise. These results highlight the relevance of addressing hardware impairments in our sensing scenario. 

    From these examples, it becomes apparent that (i) while the hardware impairments lead to small SNR penalty at the transmitter, (ii) even minor hardware impairments significantly affect the sensing receiver. Moreover, the communication receiver does not require impairment knowledge for its operation. 
    From these observations, our hypothesis is that sensing is more sensitive to the considered hardware impairments than communications, which would validate the findings from \cite{chen2023modeling}.   
\end{example} 

\section{Baseline} \label{sec:baseline}
    In this section, we derive the baseline method according to model-based benchmarks, which will later be compared with end-to-end learning approaches in Sec.~\ref{sec:results}.

    \subsection{ISAC Beamformer} \label{subsec:baseline_beamformer}
    We design the baseline for the precoding mapping in Fig.~\ref{fig:system_model}, which affects both the sensing precoder $\bmf_r$, and the communication precoder $\bmf_c$ in \eqref{eq:isac_precoder_bench}, by resorting to the {well-known} beampattern synthesis approach in {\cite{zhang2018multibeam}},\cite{analogBeamformerDesign_TSP_2017}.
    We define a uniform angular grid covering $[-\pi/2, \pi/2]$ with $\Ntheta$ grid locations $\{{\vartheta_i}\}_{i=1}^{\Ntheta}$. For a given angular interval ${\bmvarthetainterval = [\vartheta_{\min}, \vartheta_{\max}]}$, 
    we denote by ${\bbold(\bmvarthetainterval)}\in \complexset{\Ntheta}{1}$ the desired beampattern over the defined angular grid, given by $[\bbold(\bmvarthetainterval)]_i = K\mathbb{I}\{\vartheta_i \in \bmvarthetainterval\}$.
    The problem of beampattern synthesis can then be formulated as 
    \begin{equation}
        \mathop{\mathrm{min}}_{\bmf_{\text{bs}}} \lVert {\bbold(\bmvarthetainterval)} - \boldPhi_a^\trpose \bmf_{\text{bs}}  \rVert_{2}^2,
    \end{equation}
    where 
    \begin{equation} \label{eq_dict_a}
        \boldPhi_a  = [{\bma}(\theta_1) \, \ldots \, {\bma}(\theta_{\Ntheta})] \in \complexset{K}{\Ntheta}
    \end{equation}
    denotes the transmit steering matrix evaluated at the grid locations. This \ac{LS} problem
    has a simple closed-form solution {\cite{zhang2018multibeam}}
    \begin{equation}\label{eq:y_bench}
        \bmf_{\text{bs}} = (\boldPhi_a^\conj \boldPhi_a^\trpose)^{-1} \boldPhi_a^\conj {\bbold(\bmvarthetainterval)}.
    \end{equation}
    The precoder $\bmf_{\text{bs}}$ in \eqref{eq:y_bench} can be normalized to meet the required transmit power constraints. 
    The ISAC precoder $\bmf$ in \eqref{eq:Z_r} and \eqref{eq:y_c} is computed by: (i) following \eqref{eq:y_bench} with $\bmvarthetainterval=[\thetamin, \thetamax]$ and $\bmvarthetainterval=[\thetatildemin, \thetatildemax]$ to obtain $\bmf_r$ and $\bmf_c$, respectively, and (ii) combining $\bmf_r$ and $\bmf_c$ according to \eqref{eq:isac_precoder_bench}.    
    
    \subsection{Multi-target Sensing Receiver} \label{subsec:baseline_sensing_rx}
    We propose to formulate the multi-target sensing problem based on the received signal $\Yrtilde$ in \eqref{eq_Z_r2} as a sparse signal recovery problem \cite{OMP_TIT_2007} and employ the OMP algorithm \cite{Mallat1993,OMP_mmWave_2016} to solve it, which represents our model-based benchmark. 
    To construct an overcomplete dictionary for OMP, we specify an angular grid $\{\theta_i\}_{i=1}^{\Ntheta}$ and a delay grid $\{\tau_j\}_{j=1}^{\Ntau}$ depending on the region of interest for target detection (i.e., the a priori information $\{\thetamin, \thetamax, \Rmin, \Rmax\}$). Then, the spatial-domain dictionary covering angular delay grid is constructed following \eqref{eq_dict_a} and the frequency-domain dictionary covering delay grids can be constructed as
    \begin{align}  \label{eq_dict_d}
      \boldPhi_d &= [ \bmrho(\tau_1) ~ \cdots ~ \bmrho(\tau_{\Ntau})  ] \in \complexset{S}{\Ntau}~.
    \end{align}
    Using \eqref{eq_dict_a} and \eqref{eq_dict_d}, the problem of multi-target sensing based on the observation in \eqref{eq_Z_r2} becomes a sparse recovery problem
    \begin{align}\label{eq_vec_yr2}
        \Yrtilde = \sum_{i=1}^{\Ntheta} \sum_{j=1}^{\Ntau} [\SSb]_{i,j} [\boldPhi_a]_{:, i}  ([\boldPhi_d]_{:, j})^\trpose  + \bmN ~,
    \end{align}
    where $\SSb \in \complexset{\Ntheta}{\Ntau}$. Here, the goal is to estimate the $T$-sparse vector $\vecc{\SSb} \in \complexset{\Ntheta \Ntau}{1}$ under the assumption $T \ll \Ntheta \Ntau$. The baseline OMP algorithm \cite{OMP_TIT_2007,OMP_mmWave_2016,yassine2022mpnet} to solve this problem is summarized in Algorithm~\ref{alg_omp_baseline}, which will serve as a foundation to the proposed \ac{MB-ML} approaches in Sec.~\ref{sec_sens_learning}.  

From the sparse structure of \eqref{eq_vec_yr2}, we can introduce the \emph{angle-delay map} as $\boldsymbol{L}=|\boldPhi_a^H \Yrtilde^{(I)} \boldPhi_d^\conj|^2$. The targets appear as peaks in this map, provided the dictionary is matched with the true steering vector,
as discussed in relation to Fig.~\ref{fig:angle_range_map}. 
    
    \begin{algorithm}[!tb]
	\caption{{OMP for Multi-Target Sensing}}
	\label{alg_omp_baseline}
	\begin{algorithmic}[1]
	    \State \textbf{Input:} Observation $\Yrtilde$ in \eqref{eq_Z_r2}, {Angular grid $\thetagrid = \{\theta_i\}_{i=1}^{\Ntheta}$, delay grid $\taugrid = \{\tau_j\}_{j=1}^{\Ntau}$, discrete angle and range resolutions $\iotatheta, \iotaR$, and} termination threshold $\delta$. {True-false flag \textit{baseline}}.
	    \State \textbf{Output:} Set {$\Psethat$}, which contains the angle and delay estimates of multiple targets $\{ (\thetahat_{t}, \tauhat_{t})\}_{t=1}^{I}$.
     \State \textbf{Initialization:} Set $I=0$, ${\Psethat} = \varnothing$, $\boldPsi_a = \boldPsi_d = [ ~ ]$. \\
     Set the residual to $\Yrtilde^{(0)} = \Yrtilde$.\\
     {Compute dictionaries $\boldPhi_a$ and $\boldPhi_d$ according to \eqref{eq_dict_a} and \eqref{eq_dict_d}, respectively.} \\
     Compute angle-delay map $\bmL(\Yrtilde^{(I)}) = \absbigs{\boldPhi_a^H \Yrtilde^{(I)} \boldPhi_d^\conj}^2$.
     \State \textbf{while} $\max_{i,j} [\bmL(\Yrtilde^{(I)})]_{i,j} > \delta$ 
	    \Indent
	    \State \label{line_argmax} Angle-delay detection: 
     \begin{align} \label{eq:omp_argmax}
         (\ihat, \jhat) = \arg \max_{i,j} [\bmL(\Yrtilde^{(I)})]_{i,j} ~.
     \end{align}
     {
     \If{\textit{baseline}} \label{line_baseline_est} $(\thetahat_I, \tauhat_I) \gets (\theta_{\ihat}, \tau_{\jhat})$.
     \Else \Comment{Differentiable OMP} \label{line_differentiable_est}
        \begin{align}
            \admaptilde &= [\angledelaymap]_{\ihat-\iotatheta:\ihat+\iotatheta,\jhat-\iotaR:\jhat+\iotaR} ~. \label{eq:alg_mask} \\
            \admapbar_{i,j} &= \text{Softmax}(\admaptilde) ~. \label{eq:alg_softmax} \\
            \thetahat_I &= \sum_{i=\ihat-\iotatheta}^{i=\ihat+\iotatheta} [\thetagrid]_i \sum_{j=1}^{2\iotaR+1} \admapbar_{i-\ihat+\iotatheta+1,j }~, \label{eq:alg_diff_thetahat} \\
            \tauhat_I &= \sum_{j=\jhat-\iotaR}^{j=\jhat+\iotaR} [\taugrid]_j \sum_{i=1}^{2\iotatheta+1}\admapbar_{i,j-\jhat+\iotaR+1} ~. \label{eq:alg_diff_tauhat} 
        \end{align}        
     \EndIf 
        \State \label{line_update_pair} Update angle-delay pairs: ${\Psethat} \leftarrow {\Psethat} \cup \{ (\thetahat_I, \tauhat_I ) \}$.
        }
        \State \label{line_update_set} Update atom sets: 
        \begin{align}
            \boldPsi_a &\leftarrow [ \boldPsi_a ~  [\boldPhi_a]_{:, \ihat} ] ~,  \\
            \boldPsi_d &\leftarrow [ \boldPsi_d ~  [\boldPhi_d]_{:, \jhat} ] ~.
        \end{align}
	\State \label{line_update_gain} Update gain estimates: 
 \begin{align}
     \alphahatb = \arg \min_{\alphab} \normb{ \Yrtilde - \sum_{t=1}^{I+1} \alpha_t [\boldPsi_a]_{:, t} ([\boldPsi_d]_{:, t})^\trpose  }_F^2 ~.
 \end{align}
        \State Update residual: 
        \begin{align} \label{eq:update_residual}
            \Yrtilde^{(I+1)} = \Yrtilde - \sum_{t=1}^{I+1} \alphahat_t [\boldPsi_a]_{:, t} ([\boldPsi_d]_{:, t})^\trpose ~.
        \end{align}
        \State \label{line_update_I} $I = I + 1$.
     \EndIndent
     \State \textbf{end while}
	\end{algorithmic}
	\normalsize
    \end{algorithm}

    \subsection{Communication Receiver} \label{susbsec:baseline_comm_rx}
    \vspace{-0.5mm}
    Since the communication receiver has access to the \ac{CSI} $\bmkappa$ in \eqref{eq:kappa}, the received signal can be expressed as $\yc = \bmkappa \odot \bmx(\bmm) + \bmn$. Optimal decoding in this case corresponds to subcarrier-wise maximum likelihood estimation according to
    \begin{align}
        \mhat_s = \arg\min_{m_s \in \setM} \abs{[\yc]_s - [\bmkappa]_s x(m_s)}^2, \label{eq:comm_rx_est}
    \end{align}
    for $s=0,...,S-1$. In \eqref{eq:comm_rx_est}, hardware impairments described in Sec.~\ref{subsec:hardware_impairments} affect the received signal $\yc$ and the CSI $\bmkappa$, through the precoder $\bmf$. Under hardware impairments, $|\bma^\trpose(\thetat_t) \bmf|^2$ is reduced for the angular sector of interest, as shown in Fig.~\ref{fig:example_bf}, which  reduces the SNR at the communication receiver. 

\section{Model-based Learning} \label{sec:md_learning}
This section describes the two \ac{MB-ML} methods developed for multi-target ISAC: (i) \textit{dictionary learning}, which learns a dictionary of steering vectors for different angles as in \cite{mateos2022model} and is suitable for unstructured impairments, as defined in Sec.~\ref{subsec:hardware_impairments}; (ii) \textit{impairment learning}, which directly learns a parameterization of the hardware impairments and thus is suitable for structured impairments, also defined in Sec.~\ref{subsec:hardware_impairments}. 
The motivation of the proposed MB-ML approaches it to parameterize existing multi-target sensing algorithms to endow them with more degrees of freedom that can account for hardware impairments.
If structured impairments are considered, impairment learning requires less learnable parameters to optimize than dictionary learning. Our hypothesis is that impairment learning can provide faster convergence and better results by particularizing the parameters to learn.
This section also defines the loss functions and the end-to-end approach to train them, the added complexity of the proposed MB-ML approach, and how inference is performed. 

    \subsection{ISAC Beamformer} \label{subsec:md_beamformer}
    \Ac{MB-ML} follows a similar operation to \eqref{eq:y_bench} to compute the precoding vector $\bmf_r$ or $\bmf_c$, given an angular interval $\bmvarthetainterval$ (as depicted in Fig.~\ref{fig:system_model}), since the operations are already differentiable. However, there are some differences in the parameters to learn depending on the considered hardware impairments described in Sec.~\ref{subsec:hardware_impairments}:
    \begin{itemize}
        \item {\emph{Dictionary learning:} This approach considers the beamformer as a function $g_{\boldPhi_a}: \realset^2 \to \mathbbC^{K}$, mapping an angular sector $\bmvarthetainterval \in \realset^2$ to an ISAC  precoder $\bmf \in \complexset{K}{1}$ and parameterized by the transmit steering matrix $\boldPhi_a$. Dictionary learning computes $\bmf_r$ and $\bmf_c$ by means of \eqref{eq:y_bench}, regarding $\boldPhi_a \in \mathbb{C}^{K\times N_{\theta}}$ as a free learnable matrix to account for unstructured impairments. This approach hence learns $KN_{\theta}$ complex parameters. }
        \item {\emph{Impairment learning:}} The new proposed impairment learning considers instead as a free learnable parameter the vector $\bmd\in{\mathbb{R}^K}$, which represents a parameterization of the structured hardware impairments. The beamformer is regarded as a function $g_{\bmd}: \realset^2 \to \mathbbC^{K}$ parameterized by $\bmd$. The transmit steering matrix in \eqref{eq_dict_a} is computed as $\boldPhi_a(\bmd) = [\aperturbed(\theta_1;\bmd) \, \ldots \, \aperturbed(\theta_{\Ngrid};\bmd)]$, and the sensing and communication precoders are computed according to \eqref{eq:y_bench}, using $\boldPhi_a(\bmd)$ instead of $\boldPhi_a$.
        Impairment learning reduces the number of learnable parameters by taking into account the structured hardware impairments of Sec.~\ref{subsec:hardware_impairments}. Indeed, it has only $K$ real parameters, which can be several order of magnitudes less than the dictionary learning approach.
    \end{itemize}
    
    \subsection{Multi-Target Sensing Receiver}\label{sec_sens_learning}
    {Due to the nondifferentiable nature of \eqref{eq:omp_argmax} in conventional OMP (Algorithm~\ref{alg_omp_baseline}), we need to develop a differentiable version that enables end-to-end learning and backpropagation.}
    The differentiable OMP algorithm for target {angle-delay} is outlined in Algorithm~\ref{alg_omp_baseline}. We introduce differentiable operations in line~\ref{line_differentiable_est} in Algorithm~\ref{alg_omp_baseline} to enable gradient backpropagation. These operations are further described below:
    
    \begin{enumerate}
        \item \emph{Window of values around $\arg \max_{i,j} [\bmL(\Yrtilde^{(I)})]_{i,j}$:} This window is determined in \eqref{eq:alg_mask} by angle and range discrete resolutions, $\iotatheta, \iotaR$, preventing interference from other targets.\footnote{\label{fn_unfeasible_index}{Elements that correspond to unfeasible (e.g., negative) indexes are discarded.}} 
        Resolutions are defined as the minimum angle or range for which two targets are indistinguishable, and in our case they are approximated as 
        \begin{align} \label{eq:resolutions}
            \deltatheta \approx \frac{2}{K}, \quad \deltaR \approx \frac{c}{2B},  
        \end{align}
        with $B=S\Deltaf$ the bandwidth of the transmitted signal. 
        In differentiable OMP, we consider the resolutions in terms of the number of grid points of the angle-delay map. Thus, we consider
        \begin{align} \label{eq:resolutions_discrete}
            \iotatheta = \bigg\lfloor \frac{\deltatheta \Ntheta}{\thetamax-\thetamin} \bigg\rfloor, \quad \iotaR = \bigg\lfloor\frac{\deltaR \Ntau}{\Rmax - \Rmin} \bigg\rfloor,  
        \end{align} 
        where $(\thetamax-\thetamin)/\Ntheta$ and $(\Rmax-\Rmin)/\Ntau$ are the angle and range grid step sizes, respectively.
        {\item \emph{Softmax:} 
        We apply in \eqref{eq:alg_softmax} the softmax operation to the selected window, creating $\admapbar_{i,j}$, with elements summing to one. Each element in $\admapbar_{i,j}$ represents the probability of a specific angle-delay pair being the true estimate.
        \item \emph{Convex combination:} We compute the angle-delay estimate in \eqref{eq:alg_diff_thetahat}, \eqref{eq:alg_diff_tauhat} as the sum of the probabilities in $\admapbar_{i,j}$, weighted by the angle and delay values corresponding to those positions in the angle-delay map. This step, inspired by MIMO systems \cite{Lemagoarou2021} and resembling the attention mechanism \cite{Vaswani2017}, may yield an off-grid estimate.}
    \end{enumerate}
    
\begin{figure*}[tb]
        \centering 
        \includegraphics[width=0.8\textwidth]{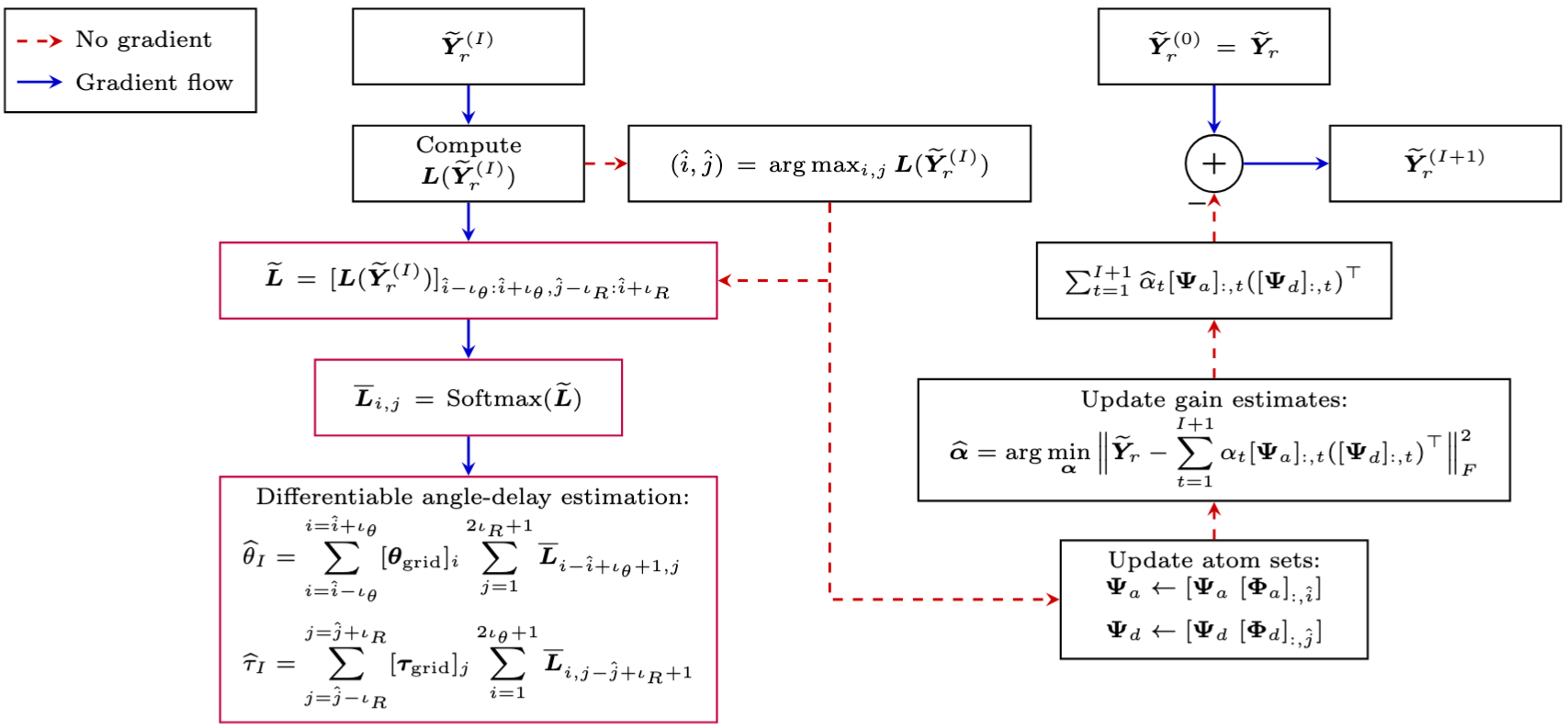}
        \caption{{Computational graph} of the $I$-th iteration of the differentiable OMP algorithm for model-based learning. Continuous {blue} lines indicate where the gradient of the loss function flows during backpropagation. {Outlined} blocks highlight modified operations with respect to conventional OMP. The algorithm stops when $\max_{i,j} [\angledelaymap]_{i,j}$ drops below a threshold.}
        \label{fig:omp_modified}
    \end{figure*}

    In differentiable OMP, we 
    utilize the same matrix $\boldPhi_a$ as the beamformer of Sec.~\ref{subsec:md_beamformer} to compute $\angledelaymap$, which allows parameter sharing between the co-located transmitter and receiver. This enables us to express the differentiable OMP algorithm as a parametric function $h_{\bmxi}: \complexset{K}{S} \to \Psethat$, where $\bmxi$ denotes $\boldPhi_a$ for dictionary learning and $\bmd$ for impairment learning. The sensing precoder is parameterized by the learnable parameters $\boldPhi_a$ and $\bmd$, which are updated based on end-to-end learning.
    {The computational graph of differentiable OMP is represented in Fig.~\ref{fig:omp_modified}, in which continuous blue lines represent backpropagation-enabled paths and dashed red lines indicate where gradient computation is disabled.} {Note that updating the atom set is based on the angle and delay dictionaries $\boldPhi_a$ and $\boldPhi_d$, respectively. This ensures that updating the residual is tied to on-grid angle-delay pairs and the same atom is never selected twice.}

    {
    \subsection{Communication Receiver} \label{sec:mb_ml_comm_rx}
    To learn the \ac{ULA} impairments with communication pilot data, we treat the message estimation problem of Sec.~\ref{subsec:comm_model} as a multi-class classification problem, as done in \cite{o2017introduction}. Therefore, we need to compute the \ac{PMF} of the transmitted messages from the received signal $\yc$ in \eqref{eq:y_c}. In our case, assuming that the transmitted messages are equiprobable, the posterior \ac{PMF} of the transmitted message for subcarrier $s$ is
    \begin{align} \label{eq:post_msg}
        p(m_s|y_s(\bmxi)) \propto \exp\bigg( -\frac{|y_s(\bmxi)-\kappa_s(\bmxi)x(m_s)|^2}{N_0S\Delta_f} \bigg)
    \end{align}
    where we denoted $[\yc]_s, [\bmkappa]_s, [\bmx(\bmm)]_s$ as $y_s, \kappa_s, x(m_s)$ for convenience. The learnable parameters $\bmxi$ affect the precoder $\bmf$ in \eqref{eq:y_c} and \eqref{eq:kappa}, which changes the SNR at the communication receiver. The PMF in \eqref{eq:post_msg} requires knowledge of the noise PSD $N_0$. To circumvent that issue, we compute the posterior PMF estimate $\bmuhat\in[0,1]^{|\setM|}$ for each subcarrier $s$ as  
    \begin{align} \label{eq:prob_comm}
        \bmuhat(\bmxi) = \text{Softmax}(-\log\abs{y_s(\bmxi) - \kappa_s(\bmxi) \bmx(\bmm)}^2)~,
    \end{align}
    where $\bmm = [1,\ldots,|\setM|]^\top$ contains all possible messages. The negative logarithmic function is applied instead of the exponential function for numerical stability. Introducing the softmax operator renders \eqref{eq:prob_comm} differentiable to perform learning.
    }
    
    \subsection{Loss Functions} \label{subsec:loss_function}
    \subsubsection{Sensing Loss Function} \label{subsubsec:sensing_loss} 
    As loss function for \ac{MB-ML}  OFDM MIMO multi-target sensing, we select the \ac{GOSPA} loss from \cite{rahmathullah2017generalized}. 
    The GOSPA loss is a well-established metric to assess the performance of multiple target tracking, which has been extensively applied in the literature \cite{pinto2023deep, jones2023gospa, wang2024dynamic}. Compared to other well-known metrics like \ac{OSPA}, GOSPA allows to penalize localization errors for detected targets and false alarm detections. This loss function also tackles the inherent data association problem between true and estimated targets.\footnote{\label{fn_GOSPA}{Note that the minimization over all permutations imposes a computational limit on the number of detectable targets. 
    However, this challenge is not unique to GOSPA but is inherent to the data association problem in multi-target tracking. 
    While there exist lower-complexity alternatives, such methods often incur a performance penalty, see, e.g., the comparison in \cite{muth2023autoencoder}. }}
    The \ac{GOSPA} loss is defined as follows. Let $\gamma>0$, $0<\mu\leq2$ and $1\leq p < \infty$. Let $\Pset = \{\bmp_1, ..., \bmp_{|\Pset|}\}$ and $\Psethat = \{\bmphat_1,...,\bmphat_{|\Psethat|}\}$ be the finite subsets of $\mathbb{R}^2$ corresponding to the true and estimated target positions, respectively, with $0\leq|\Pset|\leq\Tmax$, $0\leq|\Psethat|\leq\Tmax$. Let $d(\bmp, \bmphat) = \lVert\bmp - \bmphat\rVert_2$ be the distance between true and estimated positions, and $\dcut(\bmp, \bmphat) = \min(d(\bmp, \bmphat),\gamma)${, where $\gamma$ is} the cut-off distance. Let $\Pi_n$ be the set of all permutations of $\{1,...,n\}$ for any $n \in \mathbb{N}$ and any element $\pi \in \Pi_n$ be a sequence $(\pi(1),...,\pi(n))$. For $|\Pset| \leq |\Psethat|$, the \ac{GOSPA} loss function is defined as
    \begin{align}
        &\mathcal{J}_p^{(\gamma,\mu)}(\Pset, \Psethat;\bmxi) = \nonumber \\
        &\bigg( \min_{\pi\in \Pi_{|\Psethat|}} \sum_{i=1}^{|\Pset|} \dcut(\bmp_i, \bmphat_{\pi(i)}(\bmxi))^p + \frac{\gamma^p}{\mu}  (\abs{\Psethat}-\abs{\Pset}) \bigg)^{\frac{1}{p}}.
        \label{eq:gospa_loss}
    \end{align}
    If $\abs{\Pset} > \abs{\Psethat}, \mathcal{J}_p^{(\gamma,\mu)}(\Pset, \Psethat;\bmxi) = \mathcal{J}_p^{(\gamma,\mu)}(\Psethat, \Pset;\bmxi)$. {The larger $p$ is, the more severe penalization is applied to those estimations far from all ground-truth targets.} The value of $\gamma$ dictates the maximum allowable distance error. The role of $\mu$, together with $\gamma$, is to control the detection penalization. 
    In the GOSPA loss, the learnable parameters $\bmxi$ affect the differentiable OMP algorithm of Sec.~\ref{sec_sens_learning} as well as the precoder $\bmf$ in \eqref{eq_Z_r2}.
    Note that the minimum operation in \eqref{eq:gospa_loss} is differentiable: it selects one of the input elements and the gradient backpropagates through the selected element.

    \subsubsection{Communication Loss Function} \label{subsubsec:comm_loss_fn}
    We choose as a loss function for communications the \ac{CCE} loss \cite{o2017introduction}. A pilot message $m$ is encoded as a one-hot vector (i.e., a binary vector with only one entry as one) $\bmuenc \in \{0,1\}^{|\setM|}$. The \ac{CCE} loss between a true pilot message $\bmu_{\text{enc}}$ and a probability vector $\hat{\bmu}$ defined in \eqref{eq:prob_comm} is computed as
    \begin{align} \label{eq:comm_loss}
        \mathcal{J}_{\rm{c}}(\bmuenc, \bmuhat; \bmxi) = -\sum_{i=1}^{|\setM|} [\bmuenc]_i \log([\bmuhat(\bmxi)]_i).
    \end{align}

    \subsubsection{ISAC Loss Function} \label{subsubsec:isac_loss_fn}
    We combine the sensing and communication loss functions as a weighted sum of both losses, similarly to \cite{mateos2022end,muth2023autoencoder}, according to
    \begin{align} \label{eq:isac_loss_fn}
        \mathcal{J}_{\rm{ISAC}}(\bmxi) = \omega_r \mathcal{J}_p^{(\gamma,\mu)}(\Pset, \Psethat; \bmxi) + (1-\omega_r)\mathcal{J}_{\rm{c}}(\bmuenc, \bmuhat; \bmxi).
    \end{align}
    The value of $\omega_r$ controls the significance of the sensing loss function, with higher value implying that the ISAC system focuses more on sensing learning than communication learning.  
    The optimization problem that the proposed MB-ML approaches try to solve can be formulated as 
    \begin{equation} \label{eq:opt_problem_MBML}
        \arg\min_{\bmxi} \mathcal{J}_{\rm{ISAC}}(\bmxi),
    \end{equation}
    for a fixed value of $\omega_r$.

    \subsection{End-to-End Learning} \label{subsec:end_to_end}
    Following the block diagram in Fig.~\ref{fig:system_model}, the inputs to the ISAC system are: (i) the transmitted messages, (ii) the sensing angular sector, and (iii) the communication angular sector. As output of the sensing estimator and communication decoder, we obtain an estimate of the positions and number of targets, and the transmitted communication messages, respectively. The sensing and communication, and hence, ISAC losses of Sec.~\ref{subsec:loss_function} can be computed from these estimates. 
    To compute $\mathcal{J}_p^{(\gamma,\mu)}(\Pset, \Psethat; \bmxi)$, we share $\bmxi$ among the ISAC transceiver and sensing receiver (colored blocks in Fig.~\ref{fig:system_model}). Parameter sharing is straightforward since we consider a monostatic sensing transceiver. For $\mathcal{J}_{\rm{c}}(\bmuenc, \bmuhat; \bmxi)$, $\bmxi$ is only optimized at the ISAC transceiver. 
    Training occurs across random angular sectors for communication and sensing, enhancing inference performance over different angular ranges. At the sensing receiver, the true number of targets $T$ is presumed known. The differentiable OMP function of Sec.~\ref{sec_sens_learning} consistently estimates $T$ targets, decoupling learning from the threshold $\delta$ in Algorithm~\ref{alg_omp_baseline}. This renders the GOSPA loss without cardinality mismatch penalization term, and $\mu$ does not need to be defined.

    \subsection{Complexity} \label{subsec:complexity_mbml}
    The two proposed MB-ML approaches result in additional operations compared to standard OMP, as highlighted in Algorithm~\ref{alg_omp_baseline} and Fig.~\ref{fig:omp_modified}. In particular, the softmax operation of \eqref{eq:alg_softmax} has a complexity of $\bigO((2\iotatheta+1)(2\iotaR+1))$ and the weighted average operations of \eqref{eq:alg_diff_thetahat} and \eqref{eq:alg_diff_tauhat} have a complexity of $\bigO(2\iotatheta+1)$ and $\bigO(2\iotaR+1)$, respectively. Therefore, the added complexity of the proposed MB-ML approaches compared to standard OMP during training is approximately $\bigO((\iotatheta+1)(\iotaR+1))$ per OMP iteration. Moreover, to compute the estimated communication message in \eqref{eq:prob_comm}, the softmax operation results in an added complexity of $\bigO(|\setM|)$.
    The added complexity of backpropagation of the gradient of the loss function is negligible as only a few elements from the angle-delay map are selected in \eqref{eq:alg_mask}, reducing the amount of gradients to be computed. However, the number of learnable parameters of dictionary learning, $K\Ntheta$, is much larger than the parameters of impairment learning, $K$. Hence, we expect impairment learning to converge faster during training than dictionary learning, which will be discussed in Sec.~\ref{subsec:sensing_results}.
    During inference, we use standard beamforming, non-differentiable OMP, and communication message estimation with the learned parameters, as described in Sec.~\ref{susbsec:mbml_inference}. Standard beamforming and message estimation do not imply additional operations during inference. However, when impairments are considered in standard OMP, we no longer assume that the array is uniform and we cannot resort to efficient 2D-FFT-based algorithms to compute the angle-delay map.

    \subsection{Inference} \label{susbsec:mbml_inference}
    Once the impairments have been compensated during learning, they can be integrated in standard model-based approaches for target detection and position estimation. In our case, we freeze the learned parameters and use them for beamforming in \eqref{eq:y_bench}, the non-differentiable OMP algorithm of Sec.~\ref{subsec:baseline_sensing_rx}, and the communication receiver in \eqref{eq:comm_rx_est}. Using the non-differentiable version of the OMP algorithm also ensures that the estimated position belongs to the dictionary of considered positions in \eqref{eq_dict_a} and \eqref{eq_dict_d}, and the same atom (position) is not considered twice. The termination threshold $\delta$ can be fixed to yield a specific false alarm probability, or it can be swept over a range of values to obtain a \ac{ROC} curve. To assess the ISAC performance, we sweep over $(\eta, \phi)$ in \eqref{eq:isac_precoder_bench} at the transmitter side to obtain different precoders that distribute the radiated power among the direction of the targets or the communication receiver.
    
\section{Results} \label{sec:results}
This section details the simulation parameters and the results for multi-target \ac{ISAC}.\footnote{\label{fn_code}Source code to reproduce all numerical results in this paper will be made available at \url{https://github.com/josemateosramos/MBE2EMTISAC} after the peer-review process.}
We first describe the simulation parameters (Sec.~\ref{subsec:simulation_parameters}) and the performance metrics (Sec.~\ref{sec_perf_metric}) for communication and sensing. Then, we apply the different methods to the sensing problem (Sec.~\ref{subsec:sensing_results}), comparing the learning ability and corresponding sensing performance. This is followed by an \ac{ISAC} trade-off analysis (Sec.~\ref{subsec:isac_results}), a study on the generalization capabilities (Sec.~\ref{subsec:generalization_results}) and a comparison with standard model-based calibration (Sec.~\ref{subsec:comparison_calibration}).\footnote{\label{footnote:DL_comparison}In our results, we do not compare the proposed approaches with DL-based approaches as according to Table~\ref{tab:comparison}, \cite{mateos2022end, muth2023autoencoder, xiao2020deepfpc, wu2022doa, mateos2022model} only consider angle estimation, while in our work we consider position estimation. This implies that the designed DL architectures in \cite{mateos2022end, muth2023autoencoder, xiao2020deepfpc, wu2022doa, mateos2022model} cannot be directly applied to our work. Moreover, \cite{yassine2022mpnet, Chatelier2022} only perform channel estimation without extracting the target parameters from the channel estimates.}

\subsection{Simulation Parameters} \label{subsec:simulation_parameters}

\begin{table}[t]
\centering%
\caption{Simulation parameters}%
\label{tab:sim_parameters}%
{%
\begin{tabular}{c|c|c|c}%
\toprule%
  & Parameter & Expression & Value  \\
\midrule
\midrule
\multirow{2}{*}{\begin{tabular}[c]{@{}c@{}}Array \\ parameters\end{tabular}} & $K$ & - & 64 antennas \\
& $\bmd$ & $\normal(\lambda/2\bm{1}, \sigmalambda^2\bm{I_{K-1}})$ & $\sigmalambda = \lambda/15$  \\ \midrule
\multirow{4}{*}{\begin{tabular}[c]{@{}c@{}}OFDM \\ parameters\end{tabular}} 
&$S$ & - & 256 subcarriers \\
&$f_c$ & - & $60$ GHz  \\
&$\Deltaf$ & - & $240$ kHz  \\
&$P$ & - & $1$ W  \\ \midrule
\multirow{17}{*}{\begin{tabular}[c]{@{}c@{}}Channel \\ parameters\end{tabular}} &$\Tmax$ & -  & 5 targets  \\
&$\Tmaxtilde$ & - & 6 paths  \\
&$T$ & $\U\{0, \ldots, \Tmax\}$ & -  \\
&$\Ttilde$ & $\U\{0, \ldots, \Tmaxtilde\}$ & -  \\
&$\sigmarcst, \sigmarcstt$ & $\Exp(1/\sigmarcsmean)$  & $\sigmarcsmean = 1$ m$^2$   \\
&$\theta$ & $\U[\thetamin, \thetamax]$ & - \\
& $\thetat$ & $\U[\thetatildemin, \thetatildemax]$  & -  \\
&$\thetamin, \thetatildemin$ & $\thetamean - \Deltatheta/2$ & - \\
&$\thetamax, \thetatildemax$ & $\thetamean + \Deltatheta/2$ & - \\
&$\thetamean$ & $\U[-60\degreee, 60\degreee]$ & - \\
&$\Deltatheta$ & $\U[10\degreee, 20\degreee]$ & -  \\
&$R$ & $\U[10\ \text{m}, 43.75\ \text{m}]$ & - \\
&$\Rtilde_1$ & $\U[10\ \text{m}, 200\ \text{m}]$  & -  \\
&$\SNRr$ & $\frac{PK\Expectation[|\psit|^2]}{N_0S\Deltaf}$ & $7.05$ dB  \\
&$\SNRc$ & $\frac{1}{S}\sum_s \frac{\Expectation[|[\bmkappa]_s|^2]}{N_0S\Deltaf}$ & $7.50$ dB  \\
&$\Ntheta$ & - & $720$ \\
&$\Ntau$ & - & $200$ \\ \midrule
\multirow{6}{*}{\begin{tabular}[c]{@{}c@{}}Learning \\ parameters\end{tabular}}&$\mu$ & - & $2$ \\
&$p$ & -  & $2$ \\
&$B$ & - & $800$ samples \\
&Training iterations & - & $15,000$ \\
&\begin{tabular}[c]{@{}c@{}}Learning rate\\ (dictionary)\end{tabular} & - & $10^{-5}$ \\
&\begin{tabular}[c]{@{}c@{}}Learning rate\\ (impairment)\end{tabular} & - & $0.2$ \\
\bottomrule%
\end{tabular}%
}
\end{table}

In Table~\ref{tab:sim_parameters} we outline the considered simulation parameters of the ISAC scenario, with the following remarks: (i) for the hardware impairments $\bmd$, we consider the model of \cite{mateos2022end, rivetti2022spatial} and the antenna positions are computed as $[\bmpant]_k = [\bmpant]_0 + \sum_{i=1}^k [\bmd]_k$, $k=0,...,K-1$, with $[\bmpant]_0 = -(K-1)\lambda/4$, (ii) the mean RCS of the targets in the sensing model and the scatterers in the communication model follows a Swerling-3 model \cite[Eq.~(2.2.6)]{richards2005fundamentals}, (iii) scatterers in \eqref{eq:y_c} are distributed to ensure that there is a LOS path between transmitter and receiver and that $\Tcp$ is larger than the delay spread, i.e., $\Tcp \geq | \Rtilde_1-\Rtilde_t|/c, \forall t>1$, (iv) unless otherwise specified, the angular grid $\thetagrid$ in Algorithm~\ref{alg_omp_baseline} spans $[-\pi/2, \pi/2]$, (v) \Ac{MB-ML} is initialized with the same knowledge as the baseline, i.e., the steering vector models initially assume that $\bmd=(\lambda/2) \bm{1}$, (vi) in the \ac{GOSPA} loss, we set $\mu=2$, as recommended in \cite{rahmathullah2017generalized} and $\gamma = \infty$ during training to use all estimations during gradient backpropagation. During inference, we set $\gamma = (\Rmax-\Rmin) = 33.75$ m, which corresponds to the maximum range error, (vii) to assess the ISAC performance, we sweep $(\eta, \phi)$ in \eqref{eq:isac_precoder_bench}. We take 8 uniformly spaced values in logarithmic scale over $\eta \in [10^{-3},1]$, and we evaluate $\phi \in \{0,\pi\}$. Parameter optimization is performed using the Adam optimizer \cite{kingma2015adam}.\footnote{We tested learning rates ranging from $0.5$ to $10^{-3}$ for dictionary learning and from $10^{-3}$ to $10^{-6}$ for impairment learning.}

\subsection{Performance Metrics}\label{sec_perf_metric}
    For testing, we compute as detection performance metrics a measure of the probability of misdetection and the probability of false alarm, for multiple targets. We use the same definitions as in \cite{muth2023autoencoder}, which correspond to 
    \begin{align}
        \pmd = 1-\frac{\sum_{i=1}^B\min\{T_i, \That_i\}}{\sum_{i=1}^B T_i}, \label{eq:pmd_multi_targets}
    \end{align}
    \begin{align}
        \pfa = \frac{\sum_{i=1}^B \max\{T_i, \That_i\} - T_i}{\sum_{i=1}^B \Tmax - T_i}, \label{eq:pfa_multi_targets}
    \end{align}
    where $T_i$, $\That_i$ are the true and estimated number of targets in each batch sample, respectively. The regression performance is measured via the GOSPA loss in \eqref{eq:gospa_loss}. 

    As communication performance metric, we use the average \ac{SER} across subcarriers, computed as
    \begin{align}
        \text{SER} = \frac{1}{BS}\sum_{i=1}^B \sum_{j=1}^S \mathbb{I}\{[\bmm_i]_j \neq [\hat{\bmm}_i]_j\},
    \end{align}
    with $\bmm_i$ and $\hat{\bmm}_i$ the true and estimated message vectors at the $i$-th batch sample. 
    
    \subsection{Sensing Results} \label{subsec:sensing_results}
    Given that sensing is affected by impairments both at the transmitter and receiver, while communication is only affected at the transmitter, we first assess the sensing performance of the proposed learning approaches to evaluate their impairment compensation capabilities. We hence train both approaches purely based on sensing data (corresponding to the case of $\eta=1$ in \eqref{eq:isac_precoder_bench} and $\omega_r=1$ in \eqref{eq:isac_loss_fn}). In Fig.~\ref{fig:performance_target}, we perform a comprehensive comparison as a function of $\Tmax$ between: (i) the conventional baseline with perfect impairment knowledge, representing the lower bound in performance, (ii) the conventional baseline without knowledge of the impairments, assuming a spacing of $d=\lambda/2$, (iii) dictionary learning, and (iv) impairment learning. 
    
    \begin{figure}[tb]
        \begin{tikzpicture}[font=\scriptsize]

\begin{axis}[
width=8cm,
height=3.5cm,
legend cell align={left},
legend columns = 1,
legend style={
  fill opacity=1,
  draw opacity=1,
  text opacity=1,
  at={(0.5,1.7)},
  anchor=center
},
log basis y={10},
tick align=outside,
tick pos=left,
x grid style={white!69.0196078431373!black},
xlabel={Maximum number of targets},
xmajorgrids,
xmin=1, xmax=5,
xtick style={color=black},
y grid style={white!69.0196078431373!black},
ylabel={Misdet. probability},
ymajorgrids,
ymin=1e-2, ymax=1,
ymode=log,
ytick style={color=black},
]

\addplot[color=blue!90!black, mark=*, dashed, mark options={scale=0.8}] table {result_figures/data_sensing_perf_vs_target/benchmark_target_pmd_ideal.txt};
\addlegendentry{Baseline, known impairment model};

\addplot[color=black, mark=*, dashed, mark options={scale=0.8}] table {result_figures/data_sensing_perf_vs_target/benchmark_target_pmd_hwi.txt};
\addlegendentry{Baseline, unknown impairment model};

\addplot[color=green!40!black, mark=square, mark options={scale=0.8}] table {result_figures/data_sensing_perf_vs_target/impairment_target_pmd.txt};
\addlegendentry{Impairment learning};

\addplot[color=red!80!black, mark=square, mark options={scale=0.8}] table {result_figures/data_sensing_perf_vs_target/dictionary_target_pmd.txt};
\addlegendentry{Dictionary learning};

\end{axis}


\begin{axis}[
width=8cm,
height=3.5cm,
at={(0,-9)},      
log basis x={10},
tick align=outside,
tick pos=left,
legend columns = 1,
legend style={
  fill opacity=1,
  draw opacity=1,
  text opacity=1,
  at={(0.5,2)},
  anchor=center
},
x grid style={white!69.0196078431373!black},
xlabel={Maximum number of targets},
xmajorgrids,
xmin=1, xmax=5,
xtick style={color=black},
y grid style={white!69.0196078431373!black},
ylabel={Position GOSPA loss},
ymajorgrids,
ymin=0.3, ymax=40,
ymode=log,
ytick style={color=black},
]
\addplot[color=blue!90!black, mark=*, dashed, mark options={scale=0.8}] table {result_figures/data_sensing_perf_vs_target/benchmark_target_gospa_ideal.txt};

\addplot[color=black, mark=*, dashed, mark options={scale=0.8}] table {result_figures/data_sensing_perf_vs_target/benchmark_target_gospa_hwi.txt};

\addplot[color=green!40!black, mark=square, mark options={scale=0.8}] table {result_figures/data_sensing_perf_vs_target/impairment_target_gospa.txt};

\addplot[color=red!80!black, mark=square, mark options={scale=0.8}] table {result_figures/data_sensing_perf_vs_target/dictionary_target_gospa.txt};

\end{axis}

\end{tikzpicture}
     
        \caption{Sensing performance as a function of the maximum number of targets for a false alarm probability of $10^{-2}$ and a maximum target range of $\Rmax=43.75$ m. The sensing testing angular sector is $\{\thetamin, \thetamax\} = \{-40\degreee, -20\degreee\}$. 
        }
        \label{fig:performance_target}
    \end{figure}
    From Fig.~\ref{fig:performance_target}, it is observed that the sensing performance degrades as the number of target increases since the probability that close and distant targets appear at the same time increases. OMP is likely to estimate the closer targets first (targets with a stronger reflection), but an estimate error produces spurious lobes during the update of the residual in \eqref{eq:update_residual}, which hinders the detection and estimation of far-away targets (targets with a weaker reflection).
    Comparing the baseline and learning methods in Fig.~\ref{fig:performance_target}, the proposed impairment learning approach can obtain similar performance to the baseline with impairment knowledge for the considered cases, while dictionary learning does not yield such a large improvement compared to the baseline without impairment knowledge.
    Dictionary learning optimizes a matrix of steering vectors without any further constraint, which does not necessarily preserve the unit magnitude and linearly increasing phase of the columns of the steering matrix. 
    However, impairment learning creates the dictionary matrix from the learned antenna positions, which maintains the structure of the columns of the steering matrix.
    This indicates that by exploiting the structure of the impairments, i.e., reducing the dimensionality of the learnable parameters according to the knowledge of the impairments, impairment learning is able to converge to a better solution than dictionary learning for the considered number of training iterations. 

    In Fig.~\ref{fig:sensing_loss} we represent the GOSPA loss during training as a function of the training iteration for impairment and dictionary learning. It is observed that impairment learning converges to a lower value of the GOSPA loss than dictionary learning, which justifies the better performance of impairment learning in Fig.~\ref{fig:performance_target}. Moreover, impairment learning converges faster than dictionary learning to a plateau since impairment learning is composed of $K$ learnable parameters against the $K\Ntheta$ learnable parameters of dictionary learning.
    \begin{figure}[tb]
        \centering
        \begin{tikzpicture}[font=\scriptsize]

\begin{axis}[
width=8cm,
height=3.5cm,
legend cell align={left},
legend columns = 2,
legend style={
  fill opacity=1,
  draw opacity=1,
  text opacity=1,
  at={(0.42,1.2)},
  anchor=center
},
tick align=outside,
tick pos=left,
x grid style={white!69.0196078431373!black},
xlabel={Training iteration},
xmajorgrids,
xmin=1, xmax=1.5e4,
xtick style={color=black},
y grid style={white!69.0196078431373!black},
ylabel={GOSPA loss [m]},
ylabel style={yshift=-10pt},
ymajorgrids,
ymin=2, ymax=11,
ytick style={color=black},
]

\addplot[color=green!40!black] table {result_figures/data_loss/impairment_loss.txt};
\addlegendentry{Impairment learning};

\addplot[color=red!80!black] table {result_figures/data_loss/dictionary_loss.txt};
\addlegendentry{Dictionary learning};

\end{axis}

\end{tikzpicture}
        \caption{GOSPA loss during training as a function of the training iteration for $\Tmax=5$ targets.}
        \label{fig:sensing_loss}
    \end{figure}
    This renders impairment learning as a suitable offline learning approach to learn structured impairment models that can effectively be used in conventional algorithms. Given the results of Figs.~\ref{fig:performance_target} and \ref{fig:sensing_loss}, we will only consider the impairment learning approach henceforth.    
        
    \subsection{ISAC Trade-off Results}\label{subsec:isac_results}
    There are two main hyper-parameters that affect the ISAC trade-off during training: $\eta$ in \eqref{eq:isac_precoder_bench}, which determines how power is distributed across sensing and communication angular sectors, and $\omega_r$ in \eqref{eq:isac_loss_fn}, which directly changes the objective function to minimize. 
    Using training data with $\eta \not \in \{0,1\}$ would lead to an SNR degradation of the received signals, which produces more estimate errors and makes the learning optimization problem more challenging. Hence, in Fig.~\ref{fig:multi_target_isac}, we assess the ISAC trade-offs when training for the extreme cases of $\omega_r = 1, \eta=1$ (pure sensing training and beamforming) and $\omega_r = 0, \eta=0$ (pure communication training and beamforming) in \eqref{eq:isac_precoder_bench} and \eqref{eq:isac_loss_fn}. Any other combination of $\omega_r, \eta$ values would yield results in-between the extreme cases. To obtain ISAC trade-offs in Fig.~\ref{fig:multi_target_isac} during evaluation, we sweep over the hyper-parameters $\eta$ and $\phi$ in \eqref{eq:isac_precoder_bench}. We also show in Fig.~\ref{fig:multi_target_isac} the baseline with and without knowledge of the impairments as a reference. {We test the results in non-overlapping angular sectors for sensing and communications to magnify the ISAC trade-offs. We consider $\{\thetamin, \thetamax\} = \{-40\degreee,-20\degreee\}$ and $\{\thetatildemin, \thetatildemax\} = \{40\degreee, 60\degreee\}$. The trade-offs between sensing and communication stem from the fact that targets and the communication receiver lie in different angular sectors, and the choice of $\eta$ in \eqref{eq:isac_precoder_bench} determines the power allocated to each angular sector (also known as the \emph{subspace trade-off} \cite{liu2023seventy}}). 
    
    Comparing the baseline results (in blue and black), we observe that the impairments incur a degradation in communication performance (apart from the sensing degradation that was observed in Fig.~\ref{fig:performance_target}), as expected from the SNR degradation described in Sec.~\ref{susbsec:baseline_comm_rx}. 
    {Considering the differences between training using sensing data (green) and training using communication data (red), we notice that training with sensing data yields a large improvement in sensing performance, and both yield similar communication performance. This illustrates that sensing is more sensitive to impairments than communications, which confirms the findings of \cite{chen2023modeling} and our hypothesis of Sec.~\ref{subsec:hardware_impairments}. The sensitivity difference results from the fact that the sensing receiver exploits the phase differences across the receiver ULA to compute the angle-delay map and estimate target angles, while the communication receiver is affected by the SNR degradation at the ISAC transmitter due to impairments. As observed in Figs.~\ref{fig:example_bf} and \ref{fig:angle_range_map}, the impact of the considered impairments is minimal in the transmitter SNR, while the impact on the angle-delay map can be substantial. This fact facilitates impairment learning to converge to a better solution when trained using only sensing data, as opposed to including communication data.}

    From Fig.~\ref{fig:multi_target_isac}, we observe that impairment learning improves the ISAC trade-offs over the impairment-agnostic baseline, which shows the effectiveness of impairment learning in practical ISAC scenarios with hardware impairments.
    Surprisingly, impairment learning can outperform the baseline with knowledge of the impairments in terms of communication performance. This can be attributed to the beamforming function in \eqref{eq:y_bench}. The beampattern synthesis of \eqref{eq:y_bench} is based on an ideal beampattern response, which does not maximize the SNR over the angular sector of interest and hence, it is not optimal. Impairment learning is able to converge to an inter-antenna spacing that can also compensate for the suboptimality of the beamforming function in \eqref{eq:y_bench}.
    \begin{figure}[tb]
        \begin{tikzpicture}[font=\scriptsize]

\begin{axis}[
width=8cm,
height=3.5cm,
legend cell align={left},
legend columns = 1,
legend style={
  fill opacity=1,
  draw opacity=1,
  text opacity=1,
  at={(0.42,1.5)},
  anchor=center
},
log basis y={10},
tick align=outside,
tick pos=left,
x grid style={white!69.0196078431373!black},
xlabel={Misdetection probability},
xmajorgrids,
xmin=2e-1, xmax=1,
xmode=log,
xtick style={color=black},
y grid style={white!69.0196078431373!black},
ylabel={SER},
ymajorgrids,
ymin=3e-05, ymax=1,
ymode=log,
ytick style={color=black},
]

\addplot[color=blue!90!black, dashed, mark=*, mark options={scale=0.8}] table {result_figures/data_multi_target_isac/benchmark_pmd_ser_ideal.txt};
\addlegendentry{Baseline, known impairment model};

\addplot[color=black, dashed, mark=*, mark options={scale=0.8}] table {result_figures/data_multi_target_isac/benchmark_pmd_ser_hwi.txt};
\addlegendentry{Baseline, unknown impairment model};

\addplot[color=green!50!black, mark=triangle*, mark options={scale=0.8}] table {result_figures/data_multi_target_isac/imp_sens_pmd_ser.txt};
\addlegendentry{Impairment learning, pure sensing training ($\omega_r=1, \eta=1$)};

\addplot[color=red!80!black, mark=triangle*, mark options={scale=0.8}] table {result_figures/data_multi_target_isac/imp_comm_pmd_ser.txt};
\addlegendentry{Impairment learning, pure comm. training ($\omega_r=0, \eta=0$)};

\end{axis}


\begin{axis}[
width=8cm,
height=3.5cm,
at={(0,-7)},      
log basis y={10},
tick align=outside,
tick pos=left,
legend columns = 2,
legend style={
  fill opacity=1,
  draw opacity=1,
  text opacity=1,
  at={(0.5,1.45)},
  anchor=center
},
x grid style={white!69.0196078431373!black},
xlabel={Position GOSPA loss},
xmajorgrids,
xmin=12, xmax=32,
xmode=log,
log x ticks with fixed point,
xtick style={color=black},
y grid style={white!69.0196078431373!black},
ylabel={SER},
ymajorgrids,
ymin=3e-05, ymax=1,
ymode=log,
ytick style={color=black},
]

\addplot[color=black, dashed, mark=*, mark options={scale=0.8}] table {result_figures/data_multi_target_isac/benchmark_gospa_ser_hwi.txt};

\addplot[color=blue!90!black, dashed, mark=*, mark options={scale=0.8}] table {result_figures/data_multi_target_isac/benchmark_gospa_ser_ideal.txt};

\addplot[color=green!50!black, mark=triangle*, mark options={scale=0.8}] table {result_figures/data_multi_target_isac/imp_sens_gospa_ser.txt};

\addplot[color=red!80!black, mark=triangle*, mark options={scale=0.8}] table {result_figures/data_multi_target_isac/imp_comm_gospa_ser.txt};


\end{axis}

\end{tikzpicture}
        \caption{ISAC trade-offs, when the false alarm probability is set to $10^{-2}$, the maximum number of targets to $\Tmax = 5$ targets, and the maximum target range to $\Rmax = 43.75$ m. The sensing targets lie in the angular sector $\{\thetamin, \thetamax\} = \{-40\degreee, -20\degreee\}$ and the communication receiver lies in the angular sector $\{\thetatildemin, \thetatildemax\} = \{40\degreee, 60\degreee\}$. Only optimal Pareto points are shown.}
        \label{fig:multi_target_isac}
    \end{figure}
    
    \subsection{Generalization Results} \label{subsec:generalization_results}
    A critical feature of deep learning models is their generalization capability, i.e., the capacity to give significant performance when tested with data different from the training data. In Fig.~\ref{fig:sensing_mean_angle}, we test the proposed impairment learning approach, trained with $\omega_r=\eta=\{0,0.5,1\}$, for different target angular sector $\{\thetamin, \thetamax\}$. The angular grid $\thetagrid$ in Algorithm~\ref{alg_omp_baseline} spans $[\thetamin, \thetamax]$. We fix the angular sector span to $\Deltatheta = 20\degreee$ and change the mean of the angular sector $\thetamean$. The position error is computed from \eqref{eq:gospa_loss} assuming a known number of targets. In Fig.~\ref{fig:sensing_mean_angle}, we include angular sectors that are not included in the training data described in Sec.~\ref{subsec:simulation_parameters} (when $\thetamean > 60\degreee$). It is observed that sensing performance degrades as $\thetamean$ increases due to the scan loss of the ULA. Moreover, impairment learning with $\omega_r=\eta=1$ succeeds in generalizing to the new testing data, as the performance is still quite close to that of the baseline with knowledge of the impairments. This emphasizes the advantages of MB-ML and in particular of the proposed impairment learning to offer the performance guarantees of the associated model-based algorithms (OMP in our case). Comparing the cases of $\omega_r=\eta=1$ and $\omega_r=\eta=\{0,0.5\}$, it is observed that lower values of $\omega_r, \eta$ yield worse sensing results. This is expected as lower values of $\omega_r, \eta$ result in a high weight for the communication loss function in \eqref{eq:isac_loss_fn} and communication is less sensitive to the effect of the considered impairments, as discussed in Sec.\ref{subsec:isac_results}. 
    The gap between $\omega_r=\eta=1$ and $\omega_r=\eta=\{0,0.5\}$ is reduced for small values of $\thetamean$ since the effect of the impairments is less noticeable by the definition of the steering vector. Thus, the optimization carried out by the communication loss function suffices to yield a good sensing performance.

    \begin{figure}[tb]
        \begin{tikzpicture}[font=\scriptsize]

\begin{axis}[
width=8cm,
height=3.5cm,
legend cell align={left},
legend columns = 1,
legend style={
  fill opacity=1,
  draw opacity=1,
  text opacity=1,
  at={(0.5,1.6)},
  anchor=center
},
log basis y={10},
tick align=outside,
tick pos=left,
x grid style={white!69.0196078431373!black},
xlabel={Mean of angular sector $\thetamean$ [deg]},
xmajorgrids,
xmin=10, xmax=80,
xtick style={color=black},
y grid style={white!69.0196078431373!black},
ylabel={Misdetect. prob.},
ymajorgrids,
ymin=1e-01, ymax=1,
ymode=log,
ytick style={color=black},
]

\addplot[color=blue!90!black, dashed, mark=*, mark options={scale=0.8}] table {result_figures/data_sensing_perf_unseen/benchmark_pmd_mean_angle_ideal.txt};
\addlegendentry{Baseline, known impairment model};

\addplot[color=black, dashed, mark=*, mark options={scale=0.8}] table {result_figures/data_sensing_perf_unseen/benchmark_pmd_mean_angle_hwi.txt};
\addlegendentry{Baseline, unknown impairment model};

\addplot[color=green!50!black, mark=square, mark options={scale=0.8}] table {result_figures/data_sensing_perf_unseen/imp_pmd_mean_angle_weight_1.txt};
\addlegendentry{Impairment learning, $\omega_r=1, \eta=1$};

\addplot[color=orange!70!black, dashdotted, mark=square, mark options={scale=0.8, solid}] table {result_figures/data_sensing_perf_unseen/imp_pmd_mean_angle_weight_0.5.txt};
\addlegendentry{Impairment learning, $\omega_r=0.5, \eta=0.5$};

\addplot[color=red!70!black, dotted, mark=square, mark options={scale=0.8, solid}] table {result_figures/data_sensing_perf_unseen/imp_pmd_mean_angle_weight_0.txt};
\addlegendentry{Impairment learning, $\omega_r=0, \eta=0$};
\end{axis}

\begin{axis}[
width=8cm,
height=3.5cm,
at={(0,-2.1)},      
log basis y={10},
tick align=outside,
tick pos=left,
legend columns = 2,
legend style={
  fill opacity=1,
  draw opacity=1,
  text opacity=1,
  at={(0.5,1.45)},
  anchor=center
},
x grid style={white!69.0196078431373!black},
xlabel={Mean of angular sector $\thetamean$ [deg]},
xmajorgrids,
xmin=10, xmax=80,
xtick style={color=black},
y grid style={white!69.0196078431373!black},
ylabel={Position error [m]},
ymajorgrids,
ymin=0.5, ymax=10,
ymode=log,
ytick style={color=black},
]

\addplot[color=blue!90!black, dashed, mark=*, mark options={scale=0.8}] table {result_figures/data_sensing_perf_unseen/benchmark_pos_err_angle_ideal.txt};

\addplot[color=black, dashed, mark=*, mark options={scale=0.8}] table {result_figures/data_sensing_perf_unseen/benchmark_pos_err_angle_hwi.txt};

\addplot[color=green!50!black, mark=square, mark options={scale=0.8}] table {result_figures/data_sensing_perf_unseen/imp_pos_err_angle_weight_1.txt};

\addplot[color=orange!70!black, dashdotted, mark=square, mark options={scale=0.8, solid}] table {result_figures/data_sensing_perf_unseen/imp_pos_err_angle_weight_0.5.txt};

\addplot[color=red!70!black, dotted, mark=square, mark options={scale=0.8, solid}] table {result_figures/data_sensing_perf_unseen/imp_pos_err_angle_weight_0.txt};
\end{axis}

\end{tikzpicture}
        \caption{Sensing performance as a function of the mean of the target angular sector $\thetamean$. The false alarm probability is set to $10^{-2}$, the maximum number of targets to $\Tmax = 5$ targets, and the maximum target range to $\Rmax = 43.75$ m. The span of the target angular sector is $\Deltatheta = 20\degreee$.}
        \label{fig:sensing_mean_angle}
    \end{figure}

    \subsection{Comparison with Model-based Calibration} \label{subsec:comparison_calibration}
    The proposed impairment learning approach optimizes the inter-antenna spacing in a data-driven manner. In this section, we compare impairment learning with a model-based calibration approach to optimize the inter-antenna spacing. The details of the model-based calibration approach can be found in the Appendix. 
    In Fig.~\ref{fig:results_calibration}, we compare the \ac{ROC} curve and the position error of the baseline with full knowledge of: (i) the impairments, (ii) impairment learning, and (iii) model-based calibration. We focus on sensing results since they are more severely affected by hardware impairments. Results indicate that the proposed impairment learning approach outperforms model-based calibration.
    This stems from the fact that model-based calibration optimizes each antenna element independently, while the proposed MB-ML method optimizes all antennas spacings simultaneously. 
    \begin{figure}[tb]
        \begin{tikzpicture}[font=\scriptsize, spy using outlines={circle, magnification=2.5, connect spies}]

\usetikzlibrary{spy}

\begin{axis}[
width=8cm,
height=3.5cm,
legend cell align={left},
legend columns = 1,
legend style={
  fill opacity=1,
  draw opacity=1,
  text opacity=1,
  at={(0.5,1.5)},
  anchor=center
},
tick align=outside,
tick pos=left,
x grid style={white!69.0196078431373!black},
xlabel={False alarm rate},
xmajorgrids,
xmin=1e-4, xmax=1,
xtick style={color=black},
y grid style={white!69.0196078431373!black},
ylabel={Misdet. probability},
ymajorgrids,
ymin=0, ymax=0.6,
xmode=log,
log basis x={10},
ytick style={color=black},
]

\addplot[color=blue!90!black, mark=*, dashed, mark options={scale=0.8}] table {result_figures/data_roc/baseline_ideal_pfa_pmd.txt};
\addlegendentry{Baseline, known impairment model};


\addplot[color=green!40!black, mark=square, mark options={scale=0.8}] table {result_figures/data_roc/impairment_pfa_pmd.txt};
\addlegendentry{Impairment learning};

\addplot[color=purple!80!black, dashdotted, mark=triangle, mark options={scale=0.8, solid}] table {result_figures/data_roc/baseline_learn_pfa_pmd.txt};
\addlegendentry{Model-based calibration};

\coordinate (spypoint) at (axis cs:1e-2,0.15);
\coordinate (magnifyglass) at (axis cs:1e-1,0.35);

\end{axis}

\spy [size=1cm] on (spypoint) 
  in node[fill=white, minimum size=1.3cm, circle] at (magnifyglass);


\begin{axis}[
width=8cm,
height=3.5cm,
at={(0,-325)},      
log basis x={10},
tick align=outside,
tick pos=left,
legend columns = 1,
legend style={
  fill opacity=1,
  draw opacity=1,
  text opacity=1,
  at={(0.5,2)},
  anchor=center
},
x grid style={white!69.0196078431373!black},
xlabel={False alarm rate},
xmajorgrids,
xmin=1e-4, xmax=1,
xtick style={color=black},
y grid style={white!69.0196078431373!black},
ylabel={Position error [m]},
ymajorgrids,
ymin=0, ymax=2,
xmode=log,
ytick style={color=black},
]
\addplot[color=blue!90!black, mark=*, dashed, mark options={scale=0.8}] table {result_figures/data_roc/baseline_ideal_pfa_pos_error.txt};


\addplot[color=green!40!black, mark=square, mark options={scale=0.8}] table {result_figures/data_roc/impairment_pfa_pos_error.txt};

\addplot[color=purple!80!black, dashdotted, mark=triangle, mark options={scale=0.8, solid}] table {result_figures/data_roc/baseline_learn_pfa_pos_error.txt};

\end{axis}

\end{tikzpicture}
     
        \caption{Sensing results as a function of the false alarm rate for $\Tmax=5$ and $\Rmax=43.75$ m. The sensing testing angular sector is $\{\thetamin, \thetamax\} = \{-40\degreee, -20\degreee\}$. 
        }
        \label{fig:results_calibration}
    \end{figure}

\section{Conclusions} \label{sec:conclusions}
In this work, we considered the impact of hardware impairments in multi-target OFDM ISAC systems. We showed that even small impairments can severely degrade the sensing performance (i.e., the 
ability to detect and localize targets, due to model mismatch at the sensing receiver) and to a lesser extent the communication performance (i.e., the SNR, due to reduced beamforming capabilities at the ISAC transmitter). To address this degradation, we proposed a novel \ac{MB-ML} solution to enable end-to-end learning, under different parameterizations of the hardware impairments. As part of this solution, we develop a novel differentiable version of the widely use OMP algorithm at the sensing receiver, to support different parameterizations of the hardware impairments, and a joint ISAC loss function that weights sensing and communication losses. 
We then demonstrated that: (i) impairment learning outperforms dictionary learning by using more information about the impairment structure and reducing the dimensionality of parameters to be learned, (ii) training just based on sensing data is recommended over other training procedures since it yields the best ISAC trade-off, (iii) impairment learning is able to converge to solutions that produce better precoders than the considered beamforming function, (iv) impairment learning exhibits good generalization capabilities when tested with new data, and (v) impairment learning outperforms standard model-based calibration.

As possible extensions to this work, other loss functions can be studied, as the complexity of GOSPA quickly grows with the number of targets. Moreover, this study has focused on linear models for sensing and communication. 
{Future research can explore more complex scenarios, such as considering more than one cluster of sensing targets, where beamforming cannot just be based on the angular sectors, and where powerful baselines are not readily available.}
Furthermore, our ISAC signal design approach can be extended to the frequency domain, optimizing power allocation and modulation format across subcarriers by taking into account both sensing and communication objectives \cite{MIMO_JCAS_OFDM_TWC_2023}. This includes waterfilling \cite[Ch.~4.4.1]{goldsmith2005wireless} as a special case at the communication-optimal operation point of the ISAC system. Finally, building upon the proposed MB-ML methodology, future work can also focus on accounting for imperfections in the communication link, e.g.,~\cite{yassine2022mpnet,DL_UL_ICI_2023,PAN_DL_2023}.

\appendix[Model-based Calibration] 
\label{app:md_calibration}
    Assuming the structured impairments of Sec.~\ref{subsec:hardware_impairments} and knowledge of the steering vector model, calibration of the ULA would require solving \eqref{eq:opt_problem_MBML} over all the antenna element spacings, $\bmd\in\realset^K$. This problem is high-dimensional and non-convex, which renders the solution very computationally expensive. As an alternative, we propose a simple greedy approach, which is outlined in Algorithm~\ref{alg:mb_calibration}. The proposed calibration approach is based on collecting $M$ observations $\Yrtilde$ by placing a random number of targets $T\sim\U[1,\ldots, \Tmax]$ in front of the ULA and optimize the antenna element positions. The optimization of the antenna element positions is done at the post-processing stage after receiving the $M$ observations. We assume a spacing of $\lambda/2$ to compute the beamformer at the transmitter side that yields the observations $\{\Yrtilde\}_{m=1}^{M}$. We choose a number of spacings to try of $\Nd=100$ points and a number of observations equal to the batch size of MB-ML, i.e., $M=B$. The assumed antenna element positions are initialized with a spacing of $\lambda/2$. Without loss of generality, we assume that the first antenna position, i.e., the reference antenna position, is known.
\begin{algorithm}[!tb]
\caption{Model-based greedy calibration.}
\label{alg:mb_calibration}
\begin{algorithmic}[1]
    \State \textbf{Input:} Grid of spacings to try $\bmdgrid\in\realset^{\Nd}$. Assumed antenna element positions $\bmpant\in\realset^{K}$.
    \State \textbf{Assumption:} The first antenna element position, $[\bmpant]_1$, is the same as the first true antenna position. The number of targets in the environment is known.
    \State \textbf{Output:} Optimized inter-antenna positions $\bmpant$.
    \State \textbf{for} $i=1,\ldots,I$
    \Indent
    \State Collect observations $\{\Yrtilde\}_{m=1}^{M}$ according to \eqref{eq_Z_r2}.
    \State \textbf{for} $k=2,\ldots,K$ \Comment{The first position is known.}
    \Indent
    \State $\bmJcal \leftarrow [\ ]$
    \State \textbf{for} $m=1,\ldots,\Nd$
    \Indent
        \State $[\bmpant]_k = [\bmpant]_{k-1} + [\bmdgrid]_m$.
        \State \parbox[t]{0.8\linewidth}{ 
            Estimate the position of the target for each observation $\Yrtilde$ via standard OMP of Algorithm~\ref{alg_omp_baseline}.}
        \State \parbox[t]{0.8\linewidth}{%
            Compute $\Expectation[\mathcal{J}_p^{(\gamma,\mu)}(\Pset, \Psethat)]$ over observations following \eqref{eq:gospa_loss}.}
        \State \parbox[t]{0.8\linewidth}{%
            $\bmJcal \leftarrow [\bmJcal\ \Expectation[\mathcal{J}_p^{(\gamma,\mu)}(\Pset, \Psethat)]]$.}
    \EndIndent
    \State \textbf{end for}
    \State $\hat{m} = \arg\min_{m=1,\ldots,\Nd} [\bmJcal]_m$.
    \State $[\bmpant]_k = [\bmpant]_{k-1} + [\bmdgrid]_{\hat{m}}$.
    \EndIndent
    \State \textbf{end for}
    \EndIndent
    \State \textbf{end for}
\end{algorithmic}
\normalsize
\end{algorithm}

 \balance
\bibliographystyle{IEEEtran}
\bibliography{references}







\vfill

\end{document}